\documentclass[sigconf]{acmart}
\AtBeginDocument{%
}

\copyrightyear{2026}
\acmYear{2026}
\setcopyright{cc}
\setcctype{by-nc-nd}
\acmConference[WWW '26]{Proceedings of the ACM Web Conference 2026}{April 13--17, 2026}{Dubai, United Arab Emirates}
\acmBooktitle{Proceedings of the ACM Web Conference 2026 (WWW '26), April 13--17, 2026, Dubai, United Arab Emirates}
\acmPrice{}
\acmDOI{10.1145/3774904.3793062}
\acmISBN{979-8-4007-2307-0/2026/04}

\usepackage{amsmath,amssymb,amsfonts}

\usepackage{algorithm}
\usepackage{algpseudocode} % 或 algpseudocode
\usepackage{subcaption}   % 提供 subfigure
\usepackage{multirow}
\usepackage{array}
\usepackage{siunitx}
\usepackage{hyperref}
\usepackage{url}

\usepackage{xcolor}
\usepackage{enumitem}
\usepackage{adjustbox}
\usepackage{tabularx}
\usepackage{pdfpages}
\usepackage{pifont}
\usepackage{xspace}
\makeatletter
\DeclareRobustCommand\onedot{\futurelet\@let@token\@onedot}
\def\@onedot{\ifx\@let@token.\else.\null\fi\xspace}
\def\eg{\emph{e.g}\onedot}

\definecolor{correctgreen}{RGB}{50, 168, 82}
\definecolor{wrongred}{RGB}{226, 58, 58}
\usepackage{colortbl}
\definecolor{mygray}{gray}{0.85}
\definecolor{secondcolor}{RGB}{189,215,238}
\definecolor{firstcolor}{RGB}{255,153,153}
\newcommand{\firstcolor}[1]{\cellcolor[rgb]{1,.60,.60}{#1}}
\newcommand{\secondcolor}[1]{\cellcolor[rgb]{.741,.843,.933}{#1}}
\usepackage{makecell}
\newcommand{\cmark}{\textcolor{correctgreen}{\ding{52}}}
\newcommand{\xmark}{\textcolor{wrongred}{\ding{56}}}

\usepackage{graphicx,booktabs,array}
\newcommand{\modelicon}[1]{%
\adjincludegraphics[valign=c,keepaspectratio,max height=1.9em,max width=1.8cm]{#1}%
}
\usepackage{graphicx,tabularx,booktabs,caption}

\makeatother
\begin{document}
%% The "title" command has an optional parameter,
%% allowing the author to define a "short title" to be used in page headers.
% \title{\raisebox{-.2\baselineskip}{\includegraphics[height=1.2\baselineskip]{figs/title.png}}~~XInsight: Towards Real-World Integrative Psychological Counseling Agents}
\title{\raisebox{-.2\baselineskip}{\includegraphics[height=1.2\baselineskip]{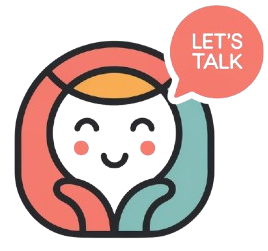}}~~XInsight: Integrative Stage-Consistent Psychological Counseling Support Agents for Digital Well-Being}
%% The "author" command and its associated commands are used to define
%% the authors and their affiliations.
%% Of note is the shared affiliation of the first two authors, and the
%% "authornote" and "authornotemark" commands
%% used to denote shared contribution to the research.
\author{Fei Wang}
% \authornote{Both authors contributed equally to this research.}
\orcid{0009-0004-1142-6434}
\affiliation{%
\institution{Hefei University of Technology}
\institution{Institute of Artificial Intelligence, Hefei Comprehensive National Science Center}
  \city{Hefei}
\country{China}
}
\email{ifei17.hfut@gmail.com}

\author{Jiangnan Yang}
\orcid{0009-0001-9028-4945}
\affiliation{
\institution{Anhui University}
\institution{Institute of Artificial Intelligence, Hefei Comprehensive National Science Center}
  \city{Hefei}
\country{China}
}
\email{jiangnanyang86@gmail.com}

\author{Junjie Chen}
\affiliation{
\institution{Hefei University of Technology}
\institution{Institute of Artificial Intelligence, Hefei Comprehensive National Science Center}
  \city{Hefei}
\country{China}
}
\orcid{0009-0001-5288-048X}
\email{jorji.chen@gmail.com}

\author{Yuxin Liu}
\orcid{0009-0003-9995-9173}
\affiliation{
\institution{Anhui University}
\institution{Institute of Artificial Intelligence, Hefei Comprehensive National Science Center}
\city{Hefei}
\country{China}
}
\email{yuxinliu221@gmail.com}

\author{Kun Li}
\authornote{Corresponding author.}
\orcid{0000-0001-5083-2145}
\affiliation{
\institution{United Arab Emirates University}
\country{AL Ain, United Arab Emirates} 
}
\email{kunli.hfut@gmail.com}

\author{Yanyan Wei}
\orcid{0000-0001-8818-6740}
\affiliation{
\institution{Hefei University of Technology}
\institution{Intelligent Interconnected Systems Laboratory of Anhui Province (HFUT)}
\city{Hefei}
\country{China}
}
\email{weiyy@hfut.edu.cn}

\author{Dan Guo}
% \authornotemark[1]
\orcid{0000-0003-2594-254X}
\affiliation{
\institution{Hefei University of Technology}
\institution{Institute of Artificial Intelligence, Hefei Comprehensive National Science Center}
%\institution{Hefei Comprehensive National Science Center}
  \city{Hefei}
\country{China}
}
\email{guodan@hfut.edu.cn}

\author{Meng Wang}
\orcid{0000-0002-3094-7735}
\affiliation{
\institution{Hefei University of Technology}
\institution{Institute of Artificial Intelligence, Hefei Comprehensive National Science Center}
  \city{Hefei}
\country{China}
}
\email{eric.mengwang@gmail.com}

\renewcommand{\shortauthors}{Fei Wang et al.}

\begin{abstract}
Web-based platforms are becoming a primary channel for psychological support, yet most LLM-driven chatbots remain opaque, single-stage, and weakly grounded in established therapeutic practice, limiting their usefulness for web applications that promote digital well-being. To address this gap, we present \textbf{XInsight}, a counseling-inspired multi-agent framework that models psychological support as a stage-consistent workflow aligned with the classical \textit{Exploration-Insight-Action} paradigm. Building on structured client representations, XInsight orchestrates specialized agents under a unified \textit{Reason-Intervene-Reflect} cycle: an Exploration agent organizes background and concerns into a structured Case Conceptualization Form, a Routing agent performs Adaptive Therapeutic Routing (ATR) across SFBT, CBT, and MBCT, a unified Therapeutic agent executes school-consistent submodules, and a Consolidation agent guides review, skill integration, and relapse-prevention planning. A Recording agent continuously transforms open-ended web dialogues into standardized psychological artifacts, including case formulations, therapeutic records, and relapse-prevention plans, enhancing interpretability, continuity, and accountability. To support rigorous and transparent assessment, we introduce \textbf{XInsight-Bench} with a Scale-Guided LLM Evaluation (SGLE) protocol that combines therapy-specific clinical scales with general counseling criteria. Experiments show improved paradigm alignment, multi-therapy integration, interaction depth, and interpretability over existing multi-agent counseling systems, indicating that XInsight provides a practical blueprint for integrating counseling-inspired support agents into web applications for digital well-being.

\end{abstract}

%%
%% The code below is generated by the tool at http://dl.acm.org/ccs.cfm.
%% Please copy and paste the code instead of the example below.
%%

%%
%% Keywords. The author(s) should pick words that accurately describe
%% the work being presented. Separate the keywords with commas.
%% A "teaser" image appears between the author and affiliation
%% information and the body of the document, and typically spans the
%% page.

% \received{20 February 2007}
% \received[revised]{12 March 2009}
% \received[accepted]{5 June 2009}

%%
%% This command processes the author and affiliation and title
%% information and builds the first part of the formatted document.
\maketitle

\begin{figure*}[t!]
\includegraphics[width=\textwidth]{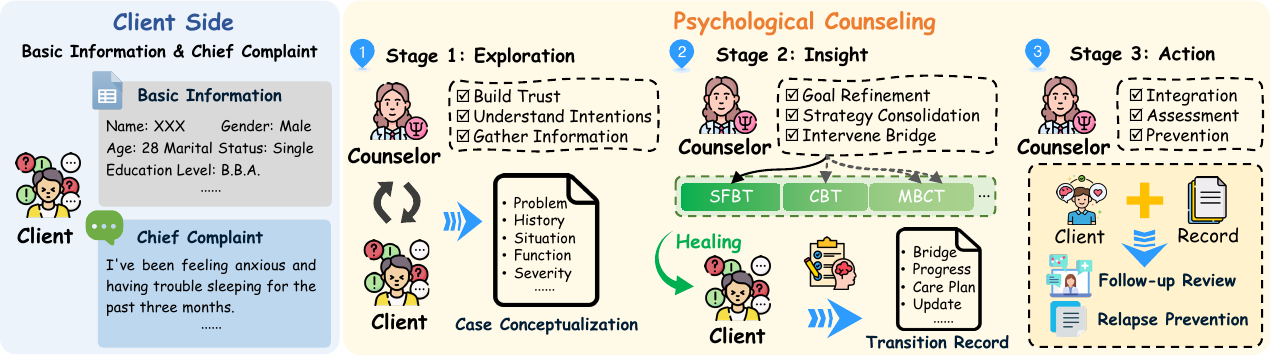}
\vspace{-2.5em}
\caption{Overview of the three-stage psychological counseling paradigm, from client input through Exploration for trust-building and case conceptualization, Insight for goal refinement and multi-school interventions, and Action for integration, assessment, and relapse prevention with follow-up review.}
\label{fig:teaser}
\vspace{-1em}
\end{figure*}
\section{Introduction}
Psychological counseling~\cite{an2025revealing,victor2022only,qiu2024interactive,chen2025mind} is a cornerstone of mental health care, helping individuals navigate complex emotional, cognitive, and behavioral difficulties~\cite{binz2023using,wang2024frequency,beck2024cognitive, kim2025mindfulness,kuyken2010does,bond2013practitioner,guo2024benchmarking,wang2025exploiting,wang2024eulermormer}.
As a growing share of support now occurs on web-based platforms and digital environments, clients increasingly seek flexible, on-demand help through online interfaces.
In practice, counseling unfolds as a dynamic, multi-stage process that progresses through \emph{Exploration}, \emph{Insight}, and \emph{Action}, rather than as a static, single-step intervention~\cite{wang2024cognition,beck2024cognitive,iftikhar2024therapy,schirmer2023increasing,li2025mmad}.
Each stage requires shifting reasoning styles, empathic attunement, and flexible selection of techniques.
Advancing systematic, stage-aware approaches is therefore important not only for extending access to mental health care but also for designing web applications that responsibly support digital well-being.

Contemporary practice commonly organizes counseling into three interlinked stages.
% , as depicted in Figure~\ref{fig:teaser}.
\textbf{\ding{182}: Exploration}, counselors and clients establish trust, clarify intentions, and gather information to construct an initial case conceptualization that reflects presenting problems, personal history, and functional impairments.
\textbf{\ding{183}: Insight}, the focus shifts toward refining goals and bridging toward change through interventions informed by Solution Focused Brief Therapy (SFBT)~\cite{franklin2012solution}, Cognitive Behavioral Therapy (CBT)~\cite{kendall2006cognitive}, and Mindfulness-Based Cognitive Therapy (MBCT)~\cite{chang2023immediate}.
\textbf{\ding{184}: Action}, counselors and clients integrate therapeutic gains, consolidate coping strategies, and plan for relapse prevention through structured follow-up and ongoing support.
Throughout, practitioners document progress via case formulations, transition records, and relapse prevention plans to maintain continuity of care~\cite{eells2022handbook,irvin1999efficacy}.

Despite rapid progress in agent-based dialogue systems~\cite{ozgun2025trustworthy,hu2025agentmental,zhu2025psi,wang2023c2d2, bond2013practitioner,zhou2025dropping}, most computational approaches remain misaligned with this staged, multi-school view of counseling and with the requirements of web-based digital mental health applications.
As summarized in Table~\ref{tab:agent}, prior systems~\cite{lee2024cactus,na2024cbt,xu2025autocbt} typically address only one of three pillars, multi-agent roles, multi-stage flow, or multi-therapy integration, and tend to sustain only short, surface-level interactions.
They struggle to follow the staged progression in Figure~\ref{fig:teaser}, dynamically adapt to evolving client needs across therapeutic schools, or ground outcomes in standardized psychological instruments.
Without such grounding, longitudinal reasoning is fragile, and evaluation often relies on dialogic heuristics rather than validated scales.

%%%%%%%%%%%%%%%%%%%%%%%%%%%%%%%%%%%%%%%%%%%%%%%%%%%%%%%%%%%%%%
\begin{table}[t!]
\caption{Comparison of representative counseling frameworks. XInsight uniquely integrates multi-agent, multi-stage, and multi-therapy capabilities, achieving full paradigm alignment and deeper interactions than prior systems.}
\vspace{-1em}
\label{tab:agent}
\centering
\renewcommand\tabcolsep{1pt}
\resizebox{1.0\linewidth}{!}{
\begin{tabular}{@{} l|ccc|cc @{}}
\toprule
\multirow{2}{*}{\textbf{Method}} & \multicolumn{3}{c|}{\textbf{Capabilities}} & \multicolumn{2}{c}{\textbf{Interactions}} \\
\cmidrule(lr){2-4}\cmidrule(l){5-6}
& \textbf{Multi-Agent} & \textbf{Mutil-Stage} & \textbf{Multi-Therapy} & \textbf{Avg.\ Turns} & \textbf{Base Model} \\ 
\midrule
\modelicon{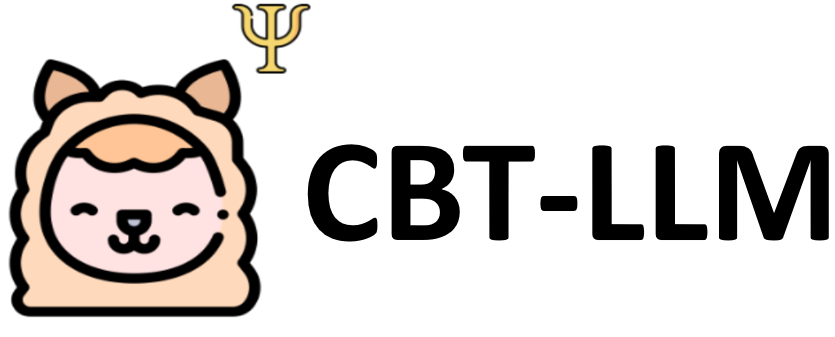}~\cite{na2024cbt}  & \xmark & \xmark & \xmark & 1.0   & GPT-3.5-Turbo  \\
\modelicon{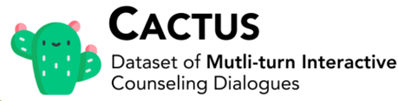}~\cite{lee2024cactus}   & \cmark & \xmark & \xmark & 16.6 & LLaMA-3-8B\\
\modelicon{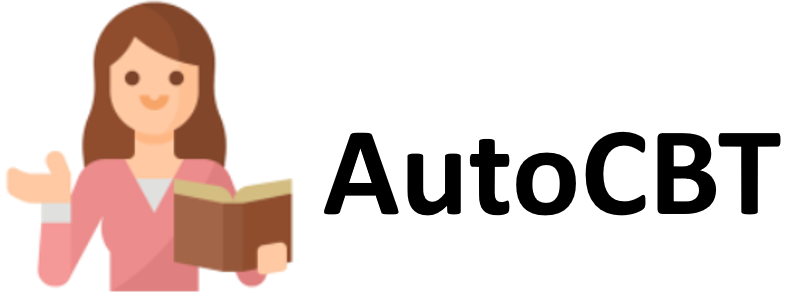}~\cite{xu2025autocbt}  & \cmark & \xmark & \xmark & 1.0    &  LLaMA-3.1-70B \\
\modelicon{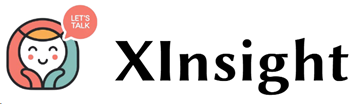} & \cmark & \cmark & \cmark & 53.7 & Qwen3-14B \\
\bottomrule
\end{tabular}
}
\vspace{-1.5em}
\end{table}

A key challenge, therefore, is how to leverage standardized tools from psychological counseling to enhance the practical applicability of multi-agent systems for web-based mental health support~\cite{chen2025mind}.
Beyond modeling the three-stage paradigm itself, this requires systematic representations of client states and counseling trajectories, together with mechanisms that can extract and formalize case formulations, stage transition records, and relapse prevention plans into interpretable and actionable data assets.
Such assets both provide contextual grounding for dynamic conversations and enable outcome evaluation and cross-therapy comparison on a reproducible basis.
There is also a pressing need for dynamic routing across heterogeneous therapeutic agents, so that the system can switch between schools adaptively while remaining consistent with the staged counseling paradigm and scalable in web environments.

%%%%%%%%%%%%%%%%%%%%%%%%%%%%%%%%%%%%%%%%%%
\begin{figure*}[t!]
\begin{center}
\includegraphics[width=1\linewidth]{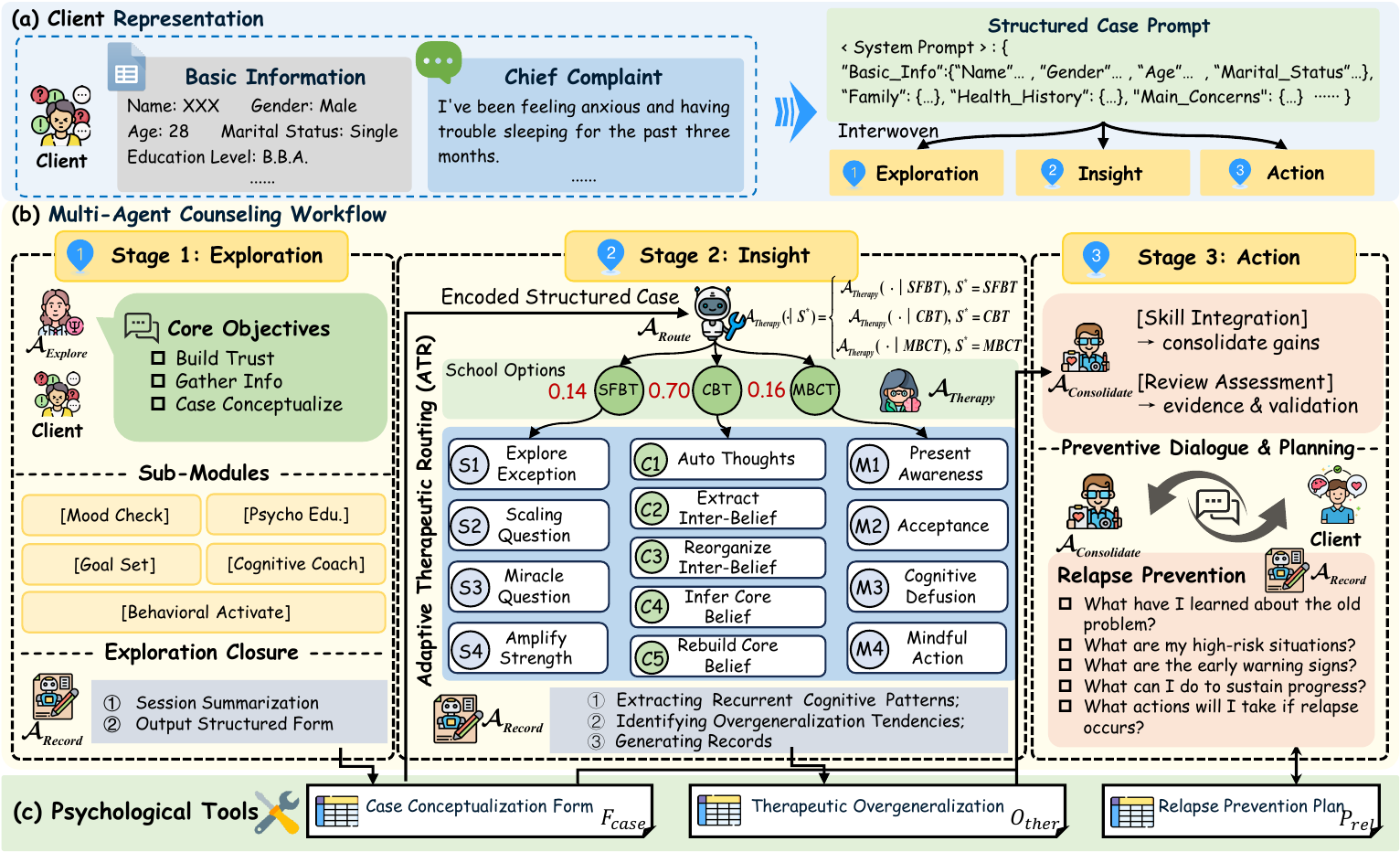}
\vspace{-2.5em}
\caption{Overview of the XInsight framework. (a) Client representation encodes basic information and the chief complaint into a structured case prompt. (b) Multi-agent counseling workflow reinterprets the three counseling stages-Exploration, Insight, and Action-supported by Adaptive Therapeutic Routing (ATR) and specialized agents. (c) Psychological tools formalize outputs such as case conceptualization, therapeutic overgeneralization, and relapse prevention.}
\label{fig:method}
\end{center}
\vspace{-1em}
\end{figure*}
%%%%%%%%%%%%%%%%%%%%%%%%%%%%%%%%%%%%%%%%%%
Motivated by these challenges, we propose \textbf{XInsight}, an integrative, paradigm-driven framework that models counseling as an evolving, stage-consistent multi-agent workflow aligned with the \emph{Exploration-Insight-Action} paradigm.
Each stage is enacted by a specialized psychological counseling support agent under a shared \textit{Reason-Intervene-Reflect} mechanism.
Agents interpret conversational context to set therapeutic intent, deliver targeted interventions consistent with SFBT, CBT, or MBCT, and distill outcomes into structured psychological documentation that informs subsequent turns.
In the \emph{Exploration} stage, the system organizes the client's background, concerns, and goals into a structured case formulation.
In the \emph{Insight} stage, an Adaptive Therapeutic Routing (ATR) procedure selects the most suitable therapeutic school, and a unified therapeutic agent executes school-consistent submodules while recording emerging evidence of change.
In the \emph{Action} stage, the system integrates prior understanding to review progress, consolidate coping skills, and design a relapse prevention plan.
Throughout, a dedicated Recording Agent applies an \emph{atomic structured tool-taking} scheme to convert open-ended web dialogues into standardized, interpretable memory artifacts (including $F_{\text{case}}$, $O_{\text{ther}}$, and $P_{\text{rel}}$), supporting continuity, transparency, and stage-consistent evaluation.
We position XInsight as a counseling-inspired support framework for web applications that promote digital well-being, designed to complement rather than replace professional mental health care.
To assess counseling quality in a way that is sensitive to therapeutic schools and robust to conversational variation, we also construct \textbf{XInsight Bench}, a cross-school benchmark with diverse client cases and a Scale Guided LLM Evaluation (SGLE) protocol that combines therapy-specific clinical scales with general counseling dimensions (see Figure~\ref{fig:bench}).
This design enables model-agnostic, reproducible assessment anchored in professional standards.

% All things considered, our contributions can be summarized as follows:
In general, the following succinctly describes our contributions:
% \vspace{-1em}
\begin{itemize}[leftmargin=*]
\item \textbf{\textit{Paradigm-Aligned Architecture.}} We recast counseling as a three-stage, multi-agent workflow governed by a unified Reason-Intervene-Reflect cycle, delivering stage-consistent control while preserving the fluidity of counseling dialogue in web settings.
\item \textbf{\textit{Adaptive Therapeutic Routing (ATR).}} We introduce an ATR procedure that selects among SFBT, CBT, and MBCT, together with a unified therapeutic agent that executes school-consistent submodules with coherent sequencing.
\item \textbf{\textit{Structured Psychological Grounding.}} We convert multi-turn conversations into standardized clinical instruments, including the Case Conceptualization Form, the Therapeutic Record, and the Relapse Prevention Plan, providing interpretable memory and enabling longitudinal reasoning.
\item \textbf{\textit{Benchmark and Evaluation.}} We collect XInsight Bench together with a Scale Guided LLM Evaluation (SGLE) protocol that combines therapy-specific clinical scales with general counseling criteria to enable reliable, clinically informed assessment, and extensively validate the effectiveness of our framework.
\end{itemize}

\section{Methodology}
\subsection{Agent Architecture}
At the heart of XInsight are five specialized agents, as shown in Figure~\ref{fig:method}, each designed to embody a core psychological function within the three-stage counseling paradigm.
Rather than functioning in isolation, these agents play complementary therapeutic roles, jointly enabling the system to reason, intervene, and reflect in a stage-consistent and psychologically coherent manner for web-based mental health support.
Their designs are grounded in observations from counseling practice, where the interplay between exploration, formulation, intervention, and consolidation is critical to client outcomes.
Against this backdrop, each agent is endowed with a distinct rationale:

\begin{itemize}[leftmargin=*]
\item \raisebox{-.4\baselineskip}{\includegraphics[height=1.3\baselineskip]{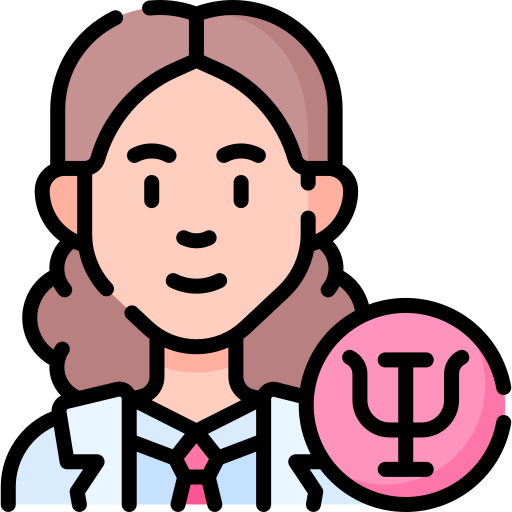}}~~\textbf{Exploration Agent ($\mathcal{A}_{\text{Explore}}$).}  
$\lozenge$~\textit{Motivation:} Early exploration in counseling relies on empathy, hypothesis generation, and flexible information gathering rather than rigid diagnostics.\\
$\lozenge$~\textit{Role:} $\mathcal{A}_{\text{Explore}}$ manages early-session dialogue to establish rapport, assess emotional states, and clarify therapeutic goals through a structured yet adaptive suite of submodules:
\emph{Mood Check} monitors affective status and symptom fluctuations;
\emph{Psycho Edu} provides corrective psychoeducation when misconceptions or stigma emerge;
\emph{Goal Set} formulates clear, achievable objectives;
\emph{Cognitive Coach} examines the content and logic of negative thoughts; and
\emph{Behavior Activate} promotes engagement in meaningful daily activities.

\item \raisebox{-.4\baselineskip}{\includegraphics[height=1.3\baselineskip]{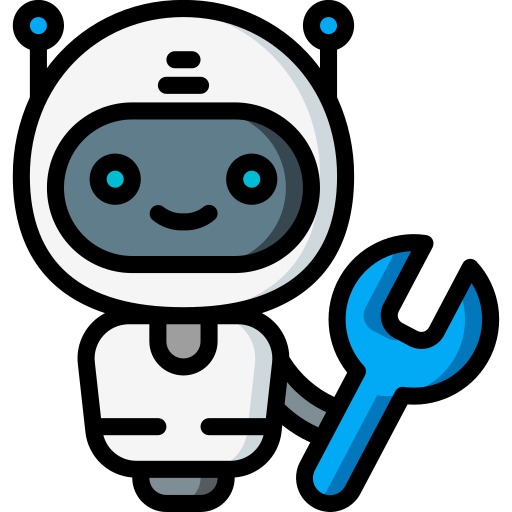}}~~\textbf{Routing Agent ($\mathcal{A}_{\text{Route}}$).} 
$\lozenge$~\textit{Motivation:} Counselors rarely adhere to a single therapeutic school, but dynamically choose approaches suited to each case formulation.
$\lozenge$~\textit{Role:} This agent implements \textbf{Adaptive Therapeutic Routing (ATR)}, analyzing the structured information in $F_{\text{case}}$ to infer the client's dominant patterns and treatment needs, and selecting the therapeutic school $S^* \in \{\text{SFBT}, \text{CBT}, \text{MBCT}\}$ that will guide the subsequent \emph{Therapeutic Agent}.
This adaptive routing keeps downstream interventions theoretically coherent, stage-consistent, and responsive to the client’s evolving state.

\item \raisebox{-.4\baselineskip}{\includegraphics[height=1.3\baselineskip]{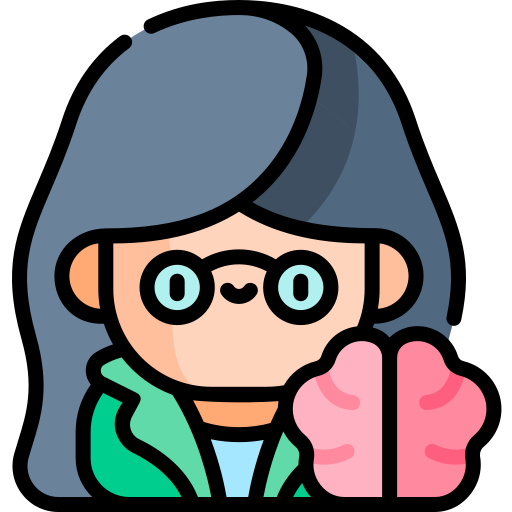}}~~\textbf{Therapeutic Agent ($\mathcal{A}_{\text{Therapy}}$).}  
$\lozenge$~\textit{Motivation:} Modern integrative therapy emphasizes flexibility, as different techniques serve the same underlying goal of insight generation and behavior change.
$\lozenge$~\textit{Role:} $\mathcal{A}_{\text{Therapy}}$ acts as a unified intervention entity whose behavior is conditioned on the therapeutic school $S^*$ selected by $\mathcal{A}_{\text{Route}}$.  
\begin{equation}
\mathcal{A}_{\text{Therapy}}(\cdot \mid S^*) =
\begin{cases}
\mathcal{A}_{\text{Therapy}}(\cdot \mid \text{SFBT}), & S^*=\text{SFBT} \\
\mathcal{A}_{\text{Therapy}}(\cdot \mid \text{CBT}), & S^*=\text{CBT}. \\
\mathcal{A}_{\text{Therapy}}(\cdot \mid \text{MBCT}), & S^*=\text{MBCT}
\end{cases}
\end{equation}
Each instantiation defines a specialized reasoning policy that sequentially selects and executes submodules.
In \textbf{SFBT}, \emph{Explore Exception} (S1), \emph{Scaling Question} (S2), \emph{Miracle Question} (S3), and \emph{Amplify Strength} (S4) promote constructive reframing and goal-oriented solution building.
In \textbf{CBT}, \emph{Auto Thoughts} (C1), \emph{Extract Inter Belief} (C2), \emph{Reorganize Inter Belief} (C3), \emph{Infer Core Belief} (C4), and \emph{Rebuild Core Belief} (C5) drive progressive cognitive restructuring from automatic thoughts to core beliefs. 
In \textbf{MBCT}, \emph{Present Awareness} (M1), \emph{Acceptance} (M2), \emph{Cognitive Defusion} (M3), and \emph{Mindful Action} (M4) cultivate meta-awareness, acceptance, and mindful action.

\item \raisebox{-.4\baselineskip}{\includegraphics[height=1.3\baselineskip]{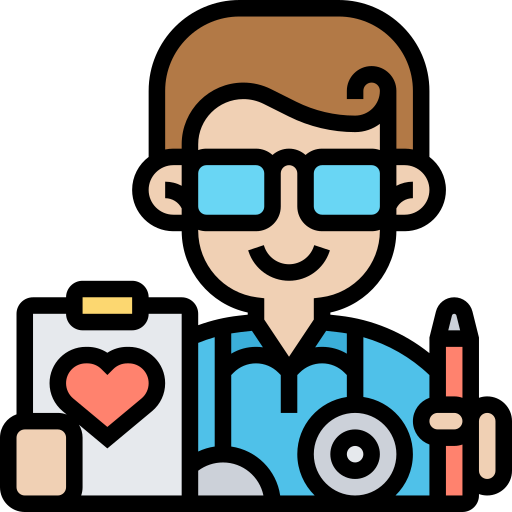}}~~\textbf{Consolidation Agent ($\mathcal{A}_{\text{Consolidate}}$).}
$\lozenge$~\textit{Motivation:} Lasting benefit depends on clients’ ability to consolidate gains and prevent relapse beyond active sessions.
$\lozenge$~\textit{Role:} $\mathcal{A}_{\text{Consolidate}}$ operationalizes this process through three submodules: \emph{Review Assessment}, \emph{Skill Integration}, and \emph{Relapse Prevention}. 
\emph{Review Assessment} guides reflection on emotional, cognitive, and behavioral changes, validating progress;
\emph{Skill Integration} supports consolidation and transfer of coping strategies to everyday contexts;
\emph{Relapse Prevention} helps identify high-risk situations, recognize early warning signs, and design maintenance plans.
During execution, $\mathcal{A}_{\text{Consolidate}}$ accesses $F_{\text{case}}$ and $O_{\text{ther}}$ as contextual anchors for integrative reflection and evidence-based planning.

\begin{figure}[t!]
\begin{center}
\includegraphics[width=1\linewidth]{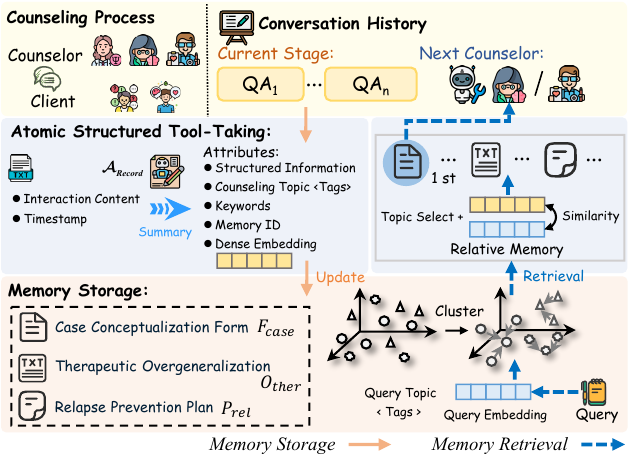}
\vspace{-2em}
\caption{Workflow of the atomic structured tool-taking mechanism in $\mathcal{A}_{\text{Record}}$, which summarizes interactions into atomic memory units, stores them as standardized psychological tools, and retrieves them via topic-guided, similarity-based search to support later agents and sessions.
}
\vspace{-2.5em}
\label{fig:mem}
\end{center}
\end{figure}

\item \raisebox{-.4\baselineskip}{\includegraphics[height=1.3\baselineskip]{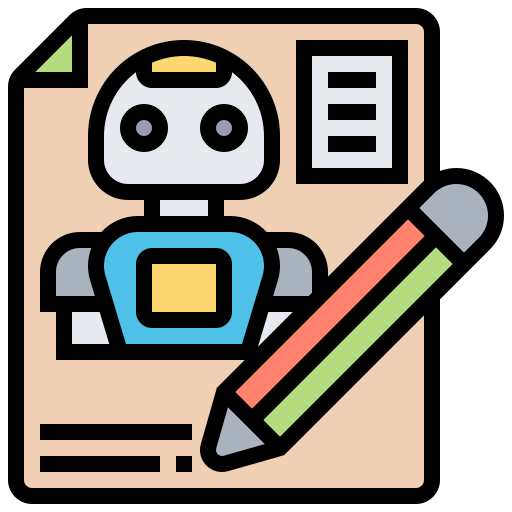}}~~\textbf{Recording Agent ($\mathcal{A}_{\text{Record}}$).} 
$\lozenge$~\textit{Motivation:} 
Yet most dialogue-based LLM systems produce unstructured text, lacking the rigor and reproducibility of formal psychological tools.
The Recording Agent addresses this gap by converting conversational content into structured, professional instruments aligned with established therapeutic frameworks and suitable for web-based mental health support.
$\lozenge$~\textit{Role:} Acting as an integrative memory and documentation layer, the Recording Agent continuously summarizes and formalizes the outputs of the \emph{Exploration}, \emph{Therapeutic}, and \emph{Consolidation} agents, grounding free-form dialogue in interpretable structure.
Following the \textbf{Atomic Structured Tool-Taking} scheme illustrated in Figure~\ref{fig:mem}, inspired by the Zettelkasten method~\cite{ahrens2022take,kadavy2021digital,xu2025mem}, traditional memory~\cite{xu2025autocbt,lee2024cactus} relies on predefined storage frameworks and is unsuitable for the multi-stage dynamic memory of counseling.
Specifically, interaction turns (Content $c_i$ and Timestamp $t_i$) are first compressed into atomic memory units annotated with structured attributes (\eg, structured information $I_i$, counseling topics $T_i$, keywords $K_i$, memory IDs $M_i$, and dense embeddings $E_i$), as:
\begin{equation}
\begin{aligned}
\{I_i, T_i, K_i\}, M_i &\leftarrow \mathcal{A}_{\text{Record}}(c_i \parallel t_i \parallel \texttt{Prompt}_{i}),\\
E_i = \phi_{enc}&(\texttt{Concat}(I_i, T_i, K_i)),
\end{aligned}
\end{equation}
where $\phi_{enc}$ denotes the text encoder and $i\in \{F_{\text{case}},O_{\text{ther}},P_{\text{rel}}\}$.
These units are then organized by $\mathcal{A}_{\text{Record}}$ into \textbf{three standardized psychological tools}: the \emph{Case Conceptualization Form} ($F_{\text{case}}$), capturing client background and presenting problems; the \emph{Therapeutic Record} ($O_{\text{ther}}$), documenting key interventions and cognitive-emotional patterns; and the \emph{Relapse Prevention Plan} ($P_{\text{rel}}$), outlining maintenance strategies and risk management.
Stored as structured long-term memory, these artifacts support topic-guided clustering and similarity-based retrieval.
Given a query $q$ from a downstream agent (with text $q_{\text{text}}$ and topic tag $\tau_q$), we first encode it into an embedding $E_q$ and restrict candidates by topic:
\begin{equation}
\begin{aligned}
E_q &= \phi_{\text{enc}}(q_{\text{text}}),~~~\mathcal{C}_{\tau_q}= \{\, i \mid \tau_q \in T_i \,\}, \\
\hat{\mathcal{M}}(q)& = \text{Top-}k\big(\operatorname{rank}\{\, \mathbb{S}(E_q, E_i) \mid i \in \mathcal{C}_{\tau_q} \,\}\big),
\end{aligned}
\label{eq:memory-retrieval}
\end{equation}
where $\mathbb{S}(\cdot,\cdot)$ denotes cosine similarity retrieval and $\hat{\mathcal{M}}(q)$ is the retrieved set of relevant memories.
As depicted in Figure~\ref{fig:mem}, this topic-filtered, similarity-based retrieval provides focused contextual cues for later stages and future sessions, enabling stage-consistent reasoning and supporting collaboration with human professionals in digital well-being applications.
\end{itemize}

\subsection{Workflow Dynamics}

As illustrated in Figure~\ref{fig:method}, the XInsight workflow progresses through a sequence of structured transformations:
\begin{equation}
D_{\text{client}} 
\xrightarrow{\mathcal{A}_{\text{Explore}}}F_{\text{case}} 
\xrightarrow{\mathcal{A}_{\text{Route}},\, \mathcal{A}_{\text{Therapy}}}O_{\text{ther}}
\xrightarrow{\mathcal{A}_{\text{Consolidate}}}P_{\text{rel}},
\end{equation}
where \(D_{\text{client}}\) denotes the client’s initial conversational input and each mapping corresponds to a counseling stage that yields both dialogue and structured artifacts.
The $\mathcal{A}_{\text{Record}}$ formalizes intermediate dialogues into interpretable psychological instruments, ensuring continuity and stage-consistent reasoning across the workflow.

\paragraph{Stage I: Exploration.}
The $\mathcal{A}_{\text{Explore}}$ agent initiates counseling by engaging in interactive dialogue with the client and providing basic information $\textit{Info}_\text{client}$:
\begin{equation}
D^{(1)} = \text{Dialogue}\big((\mathcal{A}_{\text{Explore}}, \text{Client})\mid \textit{Info}_\text{client}\big),
\end{equation}
eliciting background information, emotional states, and personal goals.
Through submodules such as \emph{Mood Check}, \emph{Psycho Edu}, \emph{Goal Set}, \emph{Cognitive Coach}, and \emph{Behavior Activate}, the agent gradually builds an understanding of the client’s biopsychosocial context.
The $\mathcal{A}_{\text{Record}}$ then transforms $D^{(1)}$ into a structured \textit{Case Conceptualization Form} $F_{\text{case}}$:
\begin{equation}
F_{\text{case}} = \mathcal{A}_{\text{Record}}(D^{(1)}),
\end{equation}
which summarizes presenting issues and preliminary therapeutic hypotheses.

\paragraph{Stage 2: Insight.}
Given \(F_{\text{case}}\), the $\mathcal{A}_{\text{Route}}$ agent performs \emph{Adaptive Therapeutic Routing (ATR)} to determine the therapeutic school:
\begin{equation}
S^* = \arg\max_{S_i \in \mathcal{S}} \, \mathbb{E}[R(S_i \mid F_{\text{case}})],
\end{equation}
where \(R(S_i)\) denotes the expected suitability of school \(S_i\) for the current formulation.
Conditioned on \(S^*\), the $\mathcal{A}_{\text{Therapy}}$ agent conducts school-specific interventions via structured interaction:
\begin{equation}
D^{(2)} = \text{Dialogue}\big((\mathcal{A}_{\text{Therapy}}(S^*), \text{Client})\mid F_{\text{case}}\big),
\end{equation}
covering, for example, solution-focused techniques, cognitive restructuring, or mindfulness practices.
These exchanges are encoded by $\mathcal{A}_{\text{Record}}$ as the \textit{Therapeutic Record} $O_{\text{ther}}$:
\begin{equation}
O_{\text{ther}} = \mathcal{A}_{\text{Record}}(D^{(2)}).
\end{equation}

\paragraph{Stage 3: Action.}
The $\mathcal{A}_{\text{Consolidate}}$ agent integrates the structured artifacts \(F_{\text{case}}\) and \(O_{\text{ther}}\) and continues dialogic reflection with the client:
\begin{equation}
D^{(3)} = \text{Dialogue}\big((\mathcal{A}_{\text{Consolidate}}, \text{Client}) \mid F_{\text{case}}, O_{\text{ther}}\big),
\end{equation}
which sequentially covers \emph{ReviewAssessment}, \emph{SkillIntegration}, and \emph{RelapsePrevention}.  
These interactions facilitate self-reflection, consolidation of coping skills, and maintenance planning.
The $\mathcal{A}_{\text{Record}}$ synthesizes these outcomes together with previous artifacts into the final \textit{Relapse Prevention Plan} $P_{\text{rel}}$:
\begin{equation}
P_{\text{rel}} = \mathcal{A}_{\text{Record}}(D^{(3)}, F_{\text{case}}, O_{\text{ther}}).
\end{equation}

Overall, XInsight establishes a transparent, stage-consistent workflow that turns client-agent dialogues into standardized psychological instruments.
The process is governed by a \textbf{Reason-Intervene-Reflect (RIR)} cycle: (i) \textit{Reason}, compute a context-aware goal and select the next submodule based on the dialogue and available artifacts; (ii) \textit{Intervene}, execute the submodule to conduct the therapeutic exchange and gather evidence; and (iii) \textit{Reflect}, update structured artifacts ($F_{\text{case}}$, $O_{\text{ther}}$, $P_{\text{rel}}$), adjust control state, and verify stage alignment.
This recurrent cycle at each stage supports interpretability, continuity, and traceable therapeutic behavior for web-based digital well-being applications.

\section{Experiments}
\subsection{Experiment Setup}
\begin{figure*}[t!]
\begin{center}
\includegraphics[width=1\linewidth]{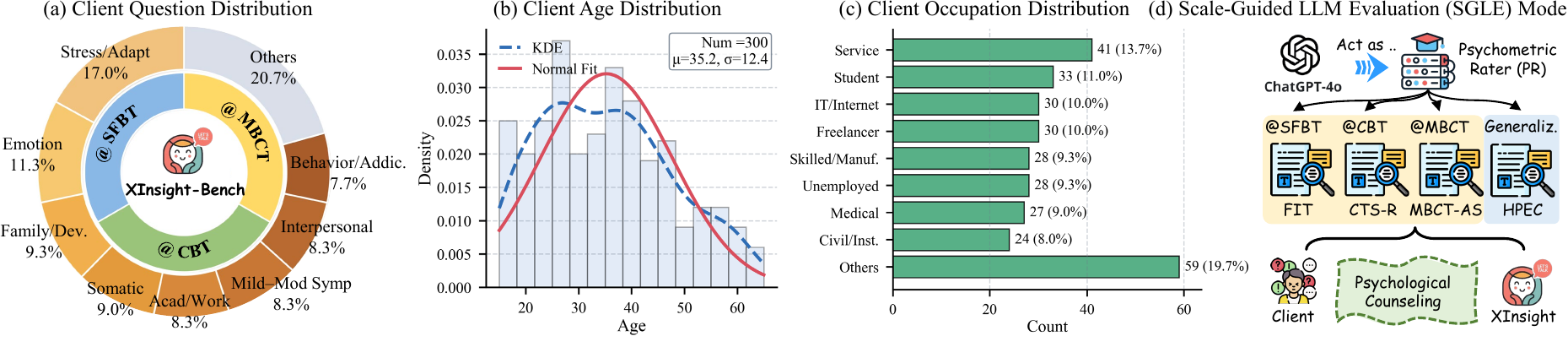}
\vspace{-2.5em}
\caption{Overview of XInsight-Bench, a benchmark designed for evaluating multi-agent psychological counseling systems. (a) Client question distribution across therapeutic schools (SFBT, CBT, MBCT) and major categories (\eg, stress/adaptation, emotion, family, somatic). (b) Client age distribution with kernel density estimation and normal fit. (c) Client occupation distribution spans service, students, IT, medical, and other groups. (d) Scale-Guided LLM Evaluation (SGLE) mode, where GPT-4o acts as a psychometric rater to score counseling quality using both school-specific standardized scales (FIT, CTS-R, MBCT-AS) and a general HPEC, enabling consistent and comprehensive evaluation across therapeutic schools.}
\label{fig:bench}
\end{center}
\vspace{-1.0em}
\end{figure*}

\noindent\textbf{XInsight-Bench.} 
 To provide a comprehensive and fair evaluation of the proposed framework, we construct \textbf{XInsight-Bench}, a benchmark tailored to multi-therapy counseling scenarios. Guided by established literature~\cite{alma2014,emilsson2014solution,beck2011cognitive}, the dataset aligns client attributes with appropriate therapeutic schools (CBT, MBCT, SFBT).
 To expand the dataset, we adopt a controlled generation strategy: GPT-4o~\cite{hurst2024gpt4o} is prompted with carefully curated exemplars to produce candidate cases, which are then filtered and reviewed by licensed counselors.
This hybrid pipeline, combining LLM-assisted generation with expert verification, yields cases that are both diverse and psychologically valid.
As shown in Figure~\ref{fig:bench}, XInsight-Bench exhibits balanced distributions across demographics and domains, providing a rigorous testbed for counseling systems (see Appendix~\ref{app_bench} for details).

%%%%%%%%%%%%%%%%%%%%%%%%%%%%%%%%%%%%%%%%%%%%%%%%%%%%%%%%%%%%%%%%
%%%%%%%%%%%%%%%%%%%%%%%%%%%%%%%%%%%%%%%%%%

%%%%%%%%%%%%%%%%%%%%%%%%%%%%%%%%%%%%%%%%%%%%%%%%%%%%%%%%%%%%%%%%  
\begin{figure}[t!]
\begin{center}
\includegraphics[width=1\linewidth]{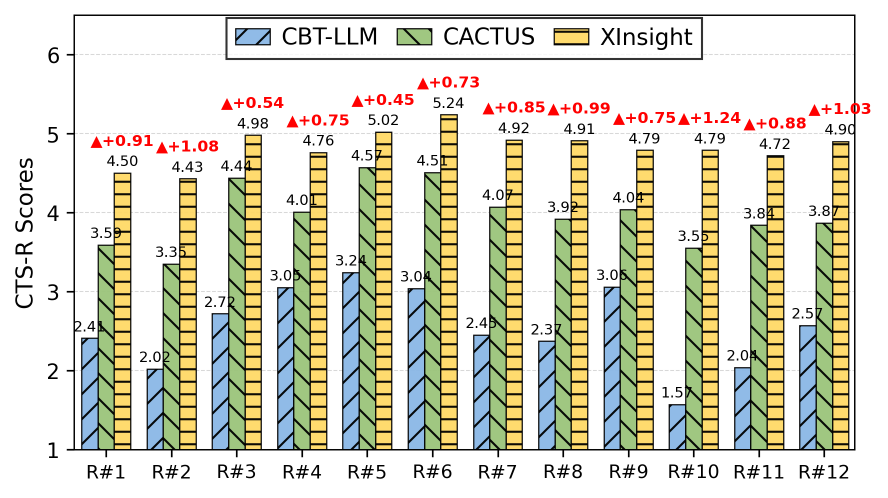}
\vspace{-2.5em}
\caption{CTS-R scores of CBT-LLM~\cite{na2024cbt}, CACTUS~\cite{lee2024cactus}, and XInsight on XInsight-Bench@CBT. 
XInsight consistently outperforms these multi-agent counseling systems.
% , as indicated by the red triangles.
}
\label{fig:sota}
\vspace{-1.8 em}
\end{center}
\end{figure}
% %%%%%%%%%%%%%%%%%%%%%%%%%%%%%%%%%%%%%%%%%%%%%%%%%%%%%%%%%%%%%%%%

\noindent\textbf{Scale-Guided LLM Evaluation.}
We design the \textbf{Scale-Guided LLM Evaluation (SGLE)} protocol (Figure~\ref{fig:bench} (d)) to benchmark counseling systems using GPT-4o. The evaluation integrates two complementary dimensions: 1) Professional Scales, employing rubrics from FIT~\cite{franklin2012solution}, CTS-R~\cite{blackburn2001revised}, and MBCT-AS~\cite{segal2002mindfulness} for standardized scoring; and 2) Human-Perspective Evaluation Criteria (HPEC), assessing attributes such as Professionalism and Safety from a client's view~\cite{lee2024cactus,yin2025mdd}. This hybrid framework ensures both psychometric rigor and scalable, objective analysis. See Appendix~\ref{app_SGLE} for details.

\noindent\textbf{Implementation Details.}
All implementations are built upon the MetaGPT framework~\cite{hong2024metagpt}, which provides a modular infrastructure for multi-agent collaboration and role specialization.
For reproducibility, the temperature parameter is fixed at 0 to ensure deterministic scoring.
To mitigate risks of recursion or stagnation in multi-stage intervention workflows, we impose structural constraints: the maximum dialogue length in the exploration stage is capped at 15 turns, and in the insight and action stages, each submodule is restricted to at most 6 consecutive selections.
We evaluate XInsight and baseline systems using several widely adopted LLM backbones as agent foundations, including InternLM2.5 (7B)~\cite{cai2024internlm2}, GLM-4 (9B)~\cite{glm2024chatglm}, Llama 3.1 (8B)~\cite{grattafiori2024llama}, Mistral (7B)~\cite{jiang2024mixtral}, Falcon-H1 (7B)~\cite{zuo2025falcon}, Gemma 3 (12B)~\cite{team2025gemma}, and Qwen2.5 (14B)~\cite{bai2023qwen}.
In subsequent experiments, we adopt Qwen3 (14B)~\cite{yang2025qwen3} as the default backbone for XInsight due to its balanced performance and efficiency.

\subsection{Main Results and Analysis}
\noindent\textbf{\ding{182} Performance across Different LLM Architectures.}
We evaluate XInsight on XInsight-Bench with a range of representative LLM backbones under three therapy-specific instruments (Table~\ref{tab:all_therapies_onecaption}): FIT for SFBT, CTS-R for CBT, and MBCT-AS for MBCT.
Results show a consistent pattern: Qwen3 (14B) attains the highest scores, reaching 74.92 on FIT, 57.96 on CTS-R, and 22.17 on MBCT-AS, surpassing strong baselines such as Qwen2.5 (14B) (69.94 / 54.53 / 19.91) and Gemma 3 (12B) (65.14 / 55.05 / 19.99).
The margin is most pronounced on MBCT-AS, where Qwen3 exceeds Qwen2.5 and Gemma 3 by 2.26 and 2.18 points, respectively, highlighting the benefit of structured paradigm alignment.
By contrast, smaller models, including InternLM2.5 (7B), Llama 3.1 (8B), and Mistral (7B), perform reasonably on CTS-R (41.43, 55.06, and 53.91), but drop markedly on FIT and MBCT-AS, indicating limited cross-school generalization.
Overall, these results suggest that paradigm-driven orchestration, rather than model size alone, is key to robust counseling quality across diverse therapeutic settings.

%%%%%%%%%%%%%%%%%%%%%%%%%%%%%%%%%%%%%%%%%%
% ====== 三表合一：一个 table*，一个 caption ======
\begin{table*}[t!]
\caption{Comparison of different LLM architectures on \textbf{XInsight-Bench} using three therapy-specific instruments: (a) FIT for SFBT, (b) CTS-R for CBT, and (c) MBCT-AS for MBCT. Each block reports per-dimension scores, with the rightmost column showing the overall sum. The best value in each column is highlighted in \colorbox{firstcolor}{red}, and the second best in \colorbox{secondcolor}{blue}.}
\label{tab:all_therapies_onecaption}
\centering
\vspace{-1em}
\fontsize{9}{11}\selectfont
\setlength{\tabcolsep}{3pt}
% \renewcommand{\arraystretch}{1.15}
% ------- (A) SFBT-FIT -------
\begin{subtable}{\linewidth}
\centering
\tabcolsep 5pt
\resizebox{\linewidth}{!}{
\begin{tabular}{l c c c c c c c c c c c c c c}
\toprule
\multirow{2}{*}{\textbf{Model (\#Params.)}}
& \multicolumn{13}{c}{\textbf{(a) @SFBT $\rightarrow$ Fidelity Instrument Therapist (FIT) (Range: 1-7, Max: 91 $\uparrow$)}} 
& \multirow{2}{*}{\makecell{\textbf{Overall$\uparrow$}\\(\textbf{Max: 91)}}} \\
\cmidrule(lr){2-14}
& \textbf{FIT\#1} & \textbf{FIT\#2} & \textbf{FIT\#3} & \textbf{FIT\#4}
& \textbf{FIT\#5} & \textbf{FIT\#6} & \textbf{FIT\#7} & \textbf{FIT\#8}
& \textbf{FIT\#9} & \textbf{FIT\#10} & \textbf{FIT\#11} & \textbf{FIT\#12} & \textbf{FIT\#13}
& \\ 
\midrule
InternLM2.5 (7B)~\cite{cai2024internlm2}
& 1.31 & 1.21 & 2.93 & 2.55 & 3.34 & 1.97 & 2.14 & 2.00 & 3.21 & 2.03 & 1.55 & 2.07 & 1.31 & 27.62 \\
GLM-4 (9B)~\cite{glm2024chatglm}
& 3.69 & 3.28 & 3.97 & 2.66 & 4.17 & 2.97 & 3.45 & 2.97 & 4.21 & 5.79 & 2.59 & 2.62 & 2.03 & 44.40 \\
Llama 3.1 (8B)~\cite{grattafiori2024llama}
& 1.45 & 1.30 & 2.35 & 3.85 & 4.45 & 3.10 & 3.40 & 2.80 & 5.60 & 5.02 & 2.50 & 2.05 & 1.45 & 39.32 \\
Mistral (7B)~\cite{jiang2024mixtral}
& 1.32 & 1.32 & 3.37 & 3.84 & 4.42 & 1.79 & 2.11 & 1.95 & 4.16 & 2.95 & 1.32 & 1.79 & 2.11 & 32.45 \\
Falcon-H1 (7B)~\cite{zuo2025falcon}
& 2.55 & 2.45 & 4.52 & 5.41 & 5.21 & 2.62 & 4.00 & 3.59 & 5.93 & 4.90 & 2.86 & 3.55 & 5.00 & 52.59 \\
Gemma 3 (12B)~\cite{team2025gemma}
& \secondcolor{4.54} & \secondcolor{4.54} & 4.96 & 4.02 & 5.71 & 3.79 & 4.32 & 4.11 & 5.89 & \secondcolor{6.21} & 3.43 & 2.46 & 4.25 & 58.23 \\
Qwen2.5 (14B)~\cite{bai2023qwen}
& 4.18 & 4.36 & \secondcolor{5.42} & \firstcolor{6.05} & \firstcolor{6.44} & \secondcolor{4.07} & \secondcolor{5.48} & \secondcolor{5.33} & \firstcolor{6.72} & 6.09 & \secondcolor{4.48} & \secondcolor{5.31} & \firstcolor{6.01} & \secondcolor{69.94} \\ 
\textbf{Qwen3 (14B)~\cite{yang2025qwen3}}
& \firstcolor{4.83} & \firstcolor{4.93} & \firstcolor{6.03} & \secondcolor{5.83} & \secondcolor{6.24} & \firstcolor{5.55} & \firstcolor{5.72} & \firstcolor{5.48} & \secondcolor{6.69} & \firstcolor{6.62} & \firstcolor{5.07} & \firstcolor{6.00} & \secondcolor{5.93} & \firstcolor{74.92} \\
\bottomrule
\end{tabular}
}
\end{subtable}
\vspace{0pt}
% ------- (B) CBT-CTS-R -------
\begin{subtable}{\linewidth}
\centering
\resizebox{\linewidth}{!}{
\begin{tabular}{l c c c c c c c c c c c c c}
\toprule
\multirow{2}{*}{\textbf{Model (\#Params.)}}
& \multicolumn{12}{c}{\textbf{(b) @CBT $\rightarrow$ Cognitive Therapy Scale-Revised (CTS-R) (Range: 1-6, Max: 72 $\uparrow$)}} 
& \multirow{2}{*}{\makecell{\textbf{Overall$\uparrow$}\\(\textbf{Max: 72)}}}\\
\cmidrule(lr){2-13}
& \textbf{CTS-R\#1} & \textbf{CTS-R\#2} & \textbf{CTS-R\#3} & \textbf{CTS-R\#4}
& \textbf{CTS-R\#5} & \textbf{CTS-R\#6} & \textbf{CTS-R\#7} & \textbf{CTS-R\#8}
& \textbf{CTS-R\#9} & \textbf{CTS-R\#10} & \textbf{CTS-R\#11} & \textbf{CTS-R\#12}
& \\ 
\midrule
InternLM2.5 (7B)~\cite{cai2024internlm2}  & 3.41 & 3.05 & 3.93 & 3.38 & 3.93 & 3.86 & 3.72 & 3.34 & 3.48 & 3.02 & 2.97 & 3.34 & 41.43 \\
GLM-4 (9B)~\cite{glm2024chatglm}  & 1.88 & 1.53 & 2.12 & 2.00 & 2.41 & 3.06 & 2.18 & 1.94 & 2.06 & 1.18 & 1.24 & 1.71 & 23.31 \\
Llama 3.1 (8B)~\cite{grattafiori2024llama}  & 3.25 & 2.75 & 3.50 & 3.20 & 3.80 & 4.05 & 3.30 & 3.25 & 3.25 & 2.65 & 2.65 & 3.20 & 38.85 \\
Mistral (7B)~\cite{jiang2024mixtral}  & \secondcolor{4.40} & 4.25 & 4.75 & \secondcolor{4.55} & 4.75 & 4.62 & 4.62 & 4.45 & 4.45 & 4.40 & 4.35 & 4.45 & 54.04 \\
Falcon-H1 (7B)~\cite{zuo2025falcon}  & 4.19 & \secondcolor{4.41} & \secondcolor{4.88} & 4.16 & \secondcolor{4.88} & 4.18 & \secondcolor{4.76} & 4.71 & 4.65 & \firstcolor{4.88} & \secondcolor{4.53} & \secondcolor{4.82} & \secondcolor{55.05} \\
Gemma 3 (12B)~\cite{team2025gemma}  & 3.34 & 2.95 & 4.01 & 3.35 & 4.27 & 4.54 & 3.89 & 4.20 & 3.87 & 2.52 & 2.96 & 3.55 & 43.45 \\
Qwen2.5 (14B)~\cite{bai2023qwen}  & 4.24 & 4.10 & 4.66 & 4.45 & 4.66 & \secondcolor{5.00} & \secondcolor{4.76} & \secondcolor{4.83} & \secondcolor{4.66} & 4.34 & 4.21 & 4.62 & 54.53 \\
\textbf{Qwen3 (14B)~\cite{yang2025qwen3}}  & \firstcolor{4.50} & \firstcolor{4.43} & \firstcolor{4.98} & \firstcolor{4.76} & \firstcolor{5.02} & \firstcolor{5.24} & \firstcolor{4.92} & \firstcolor{4.91} & \firstcolor{4.79} & \secondcolor{4.79} & \firstcolor{4.72} & \firstcolor{4.90} & \firstcolor{57.96} \\
\bottomrule
\end{tabular}
}
\end{subtable}
\vspace{0pt}
% ------- (C) MBCT-AS -------
\begin{subtable}{\linewidth}
\centering
\resizebox{\linewidth}{!}{
\begin{tabular}{l c c c c c c c c c c c c c c c c c}
\toprule
\multirow{2}{*}{\textbf{Model (\#Params.)}}
& \multicolumn{16}{c}{\textbf{(c) @MBCT $\rightarrow$ Mindfulness-Based Cognitive Therapy Adherence Scale (MBCT-AS) (Range: 1-2, Max: 32 $\uparrow$)}} 
& \multirow{2}{*}{\makecell{\textbf{Overall$\uparrow$}\\(\textbf{Max: 32)}}} \\
\cmidrule(lr){2-17}
& \textbf{AS\#1} & \textbf{AS\#2} & \textbf{AS\#3} & \textbf{AS\#4}
& \textbf{AS\#5} & \textbf{AS\#6} & \textbf{AS\#7} & \textbf{AS\#8}
& \textbf{AS\#9} & \textbf{AS\#10} & \textbf{AS\#11} & \textbf{AS\#12}
& \textbf{AS\#13} & \textbf{AS\#14} & \textbf{AS\#15} & \textbf{AS\#16}
& \\ 
\midrule
InternLM2.5 (7B)~\cite{cai2024internlm2} & 0.38 & 0.31 & 1.62 & 0.86 & 1.66 & \secondcolor{0.62} & 0.76 & 1.83 & \secondcolor{1.52} & \firstcolor{1.72} & 0.93 & 0.93 & 1.31 & 0.66 & 0.48 & 0.79 & 16.38 \\
GLM-4 (9B)~\cite{glm2024chatglm} & 0.05 & 0.03 & 0.36 & 0.08 & 0.13 & 0.03 & 0.05 & 1.90 & 0.10 & 0.10 & 0.00 & 0.97 & 1.10 & 0.49 & 0.00 & 0.00 & 5.39 \\
Llama 3.1 (8B)~\cite{grattafiori2024llama} & 0.05 & 0.11 & 0.16 & 0.13 & 0.32 & 0.05 & 0.15 & 1.55 & 0.21 & 0.12 & 0.21 & 0.11 & 0.42 & 0.21 & 0.11 & 0.11 & 4.02 \\
Mistral (7B)~\cite{jiang2024mixtral} & 0.13 & 0.10 & 1.33 & 0.57 & 1.52 & 0.27 & 0.33 & 1.80 & 1.03 & 1.43 & 0.37 & 0.47 & 1.53 & 0.60 & 0.20 & 0.27 & 11.95 \\
Falcon-H1 (7B)~\cite{zuo2025falcon} & 0.65 & \firstcolor{0.53} & 0.94 & 0.65 & \firstcolor{2.00} & 0.41 & \secondcolor{1.29} & \firstcolor{2.00} & 0.82 & 0.82 & 0.53 & 1.00 & 1.53 & 1.18 & 0.84 & 1.00 & 16.19 \\
Gemma 3 (12B)~\cite{team2025gemma}  & 0.11 & 0.21 & 1.03 & 0.37 & 1.35 & 0.12 & 0.22 & \secondcolor{1.96} & 0.18 & 0.57 & 0.29 & 1.15 & 1.57 & 0.19 & 0.12 & 0.45 & 9.89 \\
Qwen2.5 (14B)~\cite{bai2023qwen}  & \secondcolor{0.67} & 0.27 & \firstcolor{1.96} & \secondcolor{1.38} & \secondcolor{1.86} & 0.44 & 1.17 & 1.80 & 1.31 & 0.91 & \firstcolor{1.03} & \secondcolor{1.61} & \secondcolor{1.70} & \firstcolor{1.48} & \firstcolor{0.98} & \firstcolor{1.34} & \secondcolor{19.91} \\
\textbf{Qwen3 (14B)~\cite{yang2025qwen3}}  & \firstcolor{0.71} & \secondcolor{0.50} & \secondcolor{1.82} & \firstcolor{1.61} & 1.71 & \firstcolor{0.74} & \firstcolor{1.32} & \firstcolor{2.00} & \firstcolor{1.89} & \secondcolor{1.50} & \secondcolor{0.96} & \firstcolor{1.93} & \firstcolor{1.71} & \secondcolor{1.46} & \secondcolor{0.92} & \secondcolor{1.29} & \firstcolor{22.17}\\
\bottomrule
\end{tabular}
}
\end{subtable}
\vspace{-1em}
\end{table*}
%%%%%%%%%%%%%%%%%%%%%%%%%%%%%%%%%%%%%%%%%%
\noindent\textbf{\ding{183} Comparison with CBT-based Multi-Agent Frameworks.}
We further compare XInsight with representative CBT-based multi-agent counseling systems, including CBT-LLM (GPT-3.5-Turbo)~\cite{na2024cbt} and CACTUS (LLaMA-3-Instruct-8B)~\cite{lee2024cactus}.
All models are evaluated on XInsight-Bench@CBT under the same setting, with CTS-R as the primary metric.
As shown in Figure~\ref{fig:sota}, XInsight achieves consistent improvements across test cases, with gains highlighted by the red markers.
Beyond numerical advantages, these results underscore the benefit of explicitly modeling the three-stage counseling paradigm: XInsight produces outputs that are better aligned with structured CBT practice while remaining compatible with web-based multi-agent deployment.
In contrast, existing CBT-based systems exhibit limited adaptability and limited integration with therapeutic fidelity, whereas XInsight broadens the scope of multi-agent modeling while maintaining structured consistency.

\noindent\textbf{\ding{184} Human-Centered Generalization Study.}
To assess generalization beyond CBT-focused settings, we conduct a human-centered evaluation of the full XInsight-Bench across multiple therapeutic schools. As shown in Table~\ref{tab:general_eval}, models are rated along six interaction-oriented dimensions.
Compared with CBT-LLM and CACTUS, XInsight demonstrates stronger adaptability, not only within CBT but also when extended to SFBT and MBCT. Consistent gains across these dimensions indicate that XInsight can maintain high-quality counseling interactions while generalizing effectively across different therapeutic paradigms, supporting its use as a counseling-inspired support framework for digital well-being applications.

%%%%%%%%%%%%%%%%%%%%%%%%%%%%%%%%%%%%%%%%%%%%%%%%%%%%%%%%%%%%%%%%
\begin{table}[t!]
\caption{Human evaluation of psychiatric dialogue models on XInsight-Bench. 
Prof: Professionalism; Com(i)/(ii): Communication (two sub-dimensions); Flu(i)/(ii): Fluency (two sub-dimensions); Sim: Similarity; Safe: Safety (Yes:0, No:1).}
\label{tab:general_eval}
\vspace{-1em}
\centering
\tabcolsep 1pt
\resizebox{\linewidth}{!}{
\begin{tabular}{lccccccc}
\toprule
\multirow{2}{*}{\textbf{Method}} & \multicolumn{7}{c}{\textbf{Human-Perspective Evaluation Criteria (HPEC)}} \\
\cmidrule(lr){2-8}
& \textbf{Prof} $\uparrow$ & \textbf{Com(i)} $\uparrow$& \textbf{Com(ii)} $\uparrow$& \textbf{Flu(i)} $\uparrow$& \textbf{Flu(ii)} $\uparrow$& \textbf{Sim} $\uparrow$& \textbf{Safe} \\
\midrule
CBT-LLM~\cite{na2024cbt} & 6.16 & 7.16 & 6.37 & 7.68 & 8.47 & 5.95 & 0.00 \\
CACTUS~\cite{lee2024cactus} & 7.11 & 8.05 & 7.79 & 8.36 & 8.94 & 7.11 & 0.00 \\ \hline
\rowcolor{gray!20}
\textbf{XInsight} (CBT) & 8.79 & 8.95 & 8.76 & 8.68 & 8.37 & 8.42 & 0.00 \\ 
\rowcolor{gray!20}
\textbf{XInsight} (SFBT) & 7.84 & 8.47 & 8.05 & 7.37 & 5.58 & 7.37 & 0.00 \\
\rowcolor{gray!20}
\textbf{XInsight} (MBCT) & 8.32 & 8.79 & 8.47 & 7.89 & 7.11 & 8.05 & 0.00 \\
\midrule
\end{tabular}
}
\vspace{-1.5em}
\end{table}
%%%%%%%%%%%%%%%%%%%%%%%%%%%%%%%%%%%%%%%%%%%%%%%%%%%

\noindent\textbf{\ding{185} Evaluating ATR Accuracy.}
Beyond overall performance, we further assess the ATR mechanism's ability to assign case states to the correct therapeutic school. In Table~\ref{tab:CM}, ATR achieves high routing accuracy, with near-perfect performance on MBCT (F1 = 0.98) and consistently strong results on CBT (F1 = 0.90) and SFBT (F1 = 0.89).
The confusion matrix in Figure~\ref{fig:cm} shows that most errors arise between CBT and SFBT, reflecting their semantic overlap, while MBCT is almost perfectly distinguished. These findings confirm that ATR maintains therapeutic paradigms and ensures that downstream agents operate within the correct school of intervention.

\subsection{Framework Analysis}
\noindent\textbf{\ding{182} Ablation Study.}
We conduct an ablation study on XInsight-Bench@CBT under the CTS-R. All settings keep the backbone, structural constraints, and scoring protocol fixed so that differences reflect the effect of the stage design rather than model capacity.

\noindent\textbf{(1) Prompted-Only Stages.}
To test whether a single instruction can replace a stage, we collapse the entire stage into a one-sentence prompt that summarizes its intent, while leaving the other stages unchanged. Overall scores are 55.95 for Prompt Stage 1, 54.60 for Prompt Stage 2, and 55.84 for Prompt Stage 3. Relative to the full model (57.96), the drops are -2.01, -3.36, and -2.12, indicating that a single prompt cannot substitute the procedural depth and context tracking provided by the full stage logic. The largest decline occurs when simplifying Stage 2, suggesting that insight processes are especially sensitive to structured reasoning and state updates.

\noindent\textbf{(2) Removing Individual Stages.}
We further refine our approach to remove each stage entirely. Overall scores are 55.79 for \textit{w/o} Stage 1, 51.38 for \textit{w/o} Stage 2, and 55.42 for \textit{w/o} Stage 3. The gaps to the full model are -2.17, -6.58, and -2.54. Eliminating Stage 2 causes the most severe degradation, confirming its role as a bridge between exploration and action. Removing Stage 1 weakens case conceptualization and information grounding, while removing Stage 3 impairs consolidation and planning.

%%%%%%%%%%%%%%%%%%%%%%%%%%%%%%%%%%%%%%%% 测试并排
\begin{table}[t!]
\centering
\fontsize{9}{11}\selectfont
\begin{minipage}{0.55\linewidth}
\caption{ATR accuracy on XInsight-Bench across therapeutic schools.} % 添加标题
\label{tab:CM}
\vspace{-1.5em}
\centering
% \vspace{-1 em}
\begin{tabular}{l|ccc}
\toprule
\multirow{2}{*}{\textbf{School}} & \multicolumn{3}{c}{\textbf{ATR Accuracy}}  \\
& \textbf{Prec.}  % Precision
& \textbf{Recall} & \textbf{F1}  \\
\midrule
CBT  & 1.00 & 0.81 & 0.90 \\
MBCT & 0.95 & 1.00 & 0.98 \\
SFBT & 0.83 & 0.95 & 0.89 \\
\bottomrule
\end{tabular}
\vspace{-1 em}
\end{minipage}
\hfill
\begin{minipage}{0.43\linewidth}
\centering
% \vspace{-1.3 em}
\includegraphics[width=\linewidth]{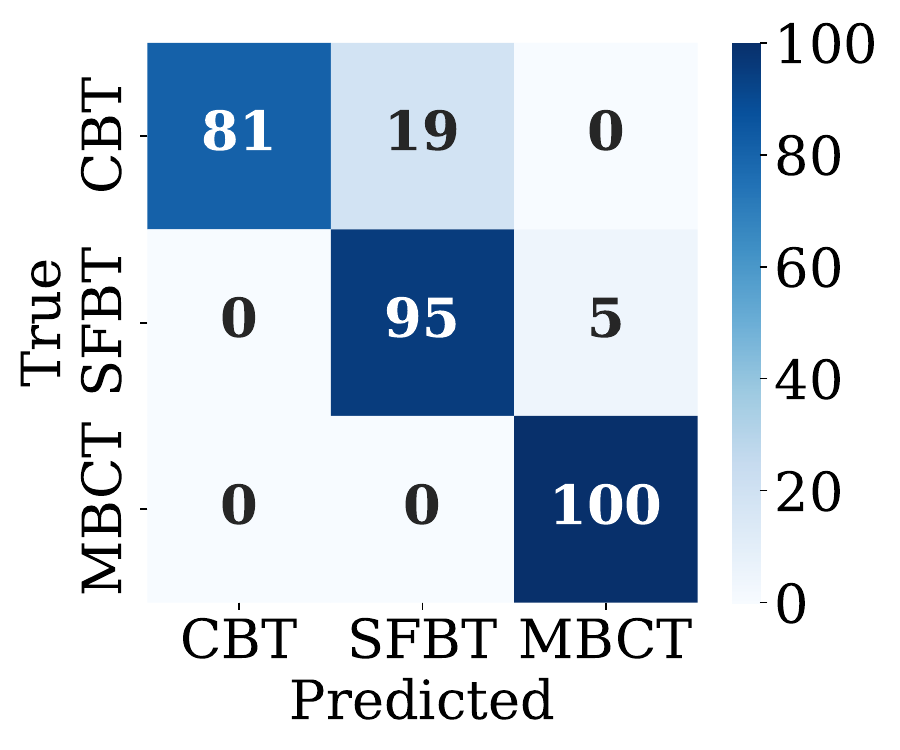}
\vspace{-2.5 em}
\captionof{figure}{Confusion matrix of therapy choice.} % 添加标题
\label{fig:cm} % 添加标签，用于引用
\end{minipage}
\vspace{-1.5em}
\end{table}
% %%%%%%%%%%%%%%%%%%%%%%%%%%%%%%%%%%%%%%%%%%%%%%%%%%%%%%%%%%%%%%%%

\begin{table*}[t!]
\caption{Ablation results on XInsight-Bench@CBT using CTS-R scores (range 1–6, max = 72). (1) Prompt Stage 1, 2, and 3 to test whether a single instruction can substitute each stage. (2) Remove Stage 1, 2, and 3 in turn to assess their individual contributions. (3) Ablate the Recording Agent and compare with a MemGPT-style memory design~\cite{packer2023memgpt}.}
\label{tab:ab}
\vspace{-1em}
\centering
\resizebox{1.0\linewidth}{!}{
\begin{tabular}{ll*{12}{c}c}
\toprule
& \multirow{2}{*}{\textbf{Setting}}
& \multicolumn{12}{c}{\textbf{Ablation Studies on @CBT $\rightarrow$ Cognitive Therapy Scale-Revised (CTS-R) (Range: 1-6, Max: 72 $\uparrow$)}} 
& \multirow{2}{*}{\makecell{\textbf{Overall$\uparrow$}\\(\textbf{Max: 72)}}} \\
\cmidrule(lr){3-14}
& & \textbf{CTS-R\#1} & \textbf{CTS-R\#2} & \textbf{CTS-R\#3} & \textbf{CTS-R\#4}
& \textbf{CTS-R\#5} & \textbf{CTS-R\#6} & \textbf{CTS-R\#7} & \textbf{CTS-R\#8}
& \textbf{CTS-R\#9} & \textbf{CTS-R\#10} & \textbf{CTS-R\#11} & \textbf{CTS-R\#12}
& \\ 
\midrule
\multirow{3}{*}{\textbf{(1)}} & Prompt Stage 1  & 4.39 & \secondcolor{4.41} & \secondcolor{4.79} & 4.63 & 4.83 & 4.93 & 4.76 & 4.76 & 4.76 & 4.45 & 4.45 & \secondcolor{4.79} & \secondcolor{55.95} \\
&Prompt Stage 2   & 4.33 & 4.11 & 4.78 & 4.56 & 4.78 & 4.89 & 4.67 & 4.62 & 4.56 & 4.43 & 4.31 & 4.56 & 54.60 \\
&Prompt Stage 3 & \firstcolor{4.58} & 4.33 & 4.75 & 4.67 & 4.75 & 4.92 & 4.67 & 4.75 & 4.67 & 4.58 & 4.50 & 4.67 & 55.84 \\ \hline
\multirow{3}{*}{\textbf{(2)}}&w/o Stage 1    & 4.43 & 4.37 & 4.63 & 4.53 & 4.68 & \secondcolor{5.16} & \secondcolor{4.79} & \secondcolor{4.79} & \secondcolor{4.74} & 4.51 & \secondcolor{4.53} & 4.63 & 55.79 \\
&w/o Stage 2     & 4.14 & 3.79 & 4.36 & 4.29 & 4.36 & 4.86 & 4.43 & 4.57 & 4.43 & 3.93 & 3.93 & 4.29 & 51.38 \\
&w/o Stage 3   & 4.45 & 4.26 & 4.71 & 4.63 & 4.71 & 5.05 & 4.74 & 4.76 & 4.61 & 4.50 & 4.37 & 4.63 & 55.42 \\\hline
\multirow{3}{*}{\textbf{(3)}}&w/o $\mathcal{A}_{\text{Record}}$   & 3.68 & 3.87 & 4.02 & 4.21 & 4.17 & 4.55 & 4.37 & 4.11 & 4.41 & 4.32 & 4.34 & 4.06 & 50.11 \\
& MemGPT~\cite{packer2023memgpt}& 4.48 & 4.29 & 4.55 & \secondcolor{4.68} & \secondcolor{4.96} & 5.05 & 4.63 & 4.75 & 4.51 & \secondcolor{4.64} & 4.30 & 4.71 & 55.55 \\
& \textbf{XInsight (Full)}  & \secondcolor{4.50} & \firstcolor{4.43} & \firstcolor{4.98} & \firstcolor{4.76} & \firstcolor{5.02} & \firstcolor{5.24} & \firstcolor{4.92} & \firstcolor{4.91} & \firstcolor{4.79} & \firstcolor{4.79} & \firstcolor{4.72} & \firstcolor{4.90} & \firstcolor{57.96} \\
\bottomrule
\end{tabular}}
\vspace{-1 em}
\end{table*}

%%%%%%%%%%%%%%%%%%%%%%%%%%%%%%%%%%%%%%%%%%%%%%%%%%%%%%%%%%%%%%%%

%%%%%%%%%%%%%%%%%%%%%%%%%%%%%%%%%%
\begin{figure*}[t!]
\begin{center}
\includegraphics[width=1\linewidth]{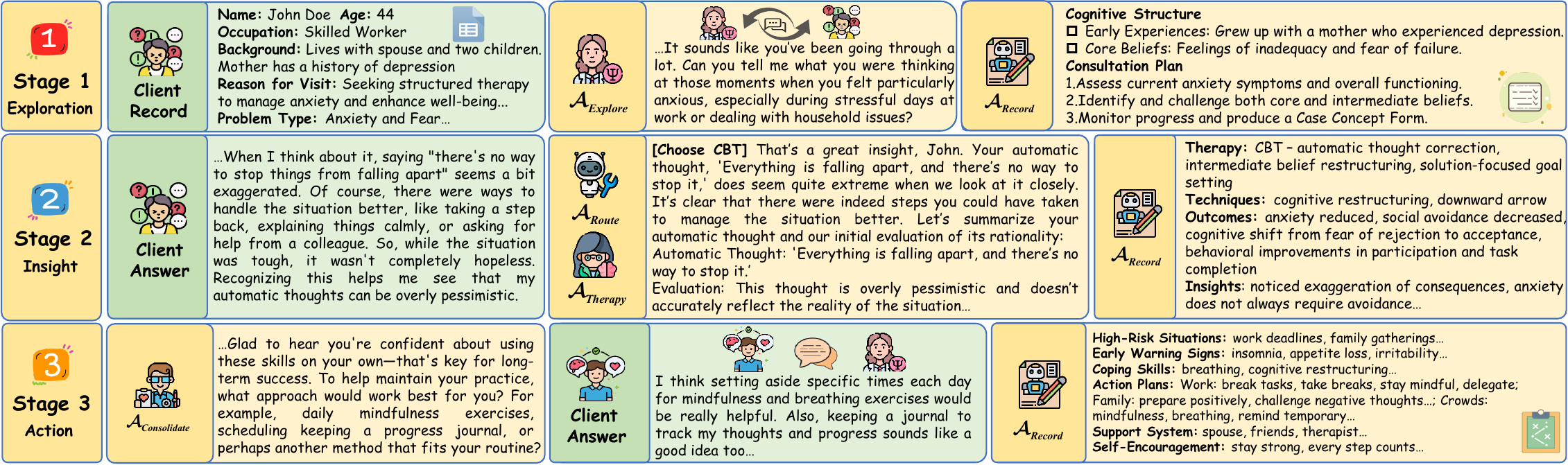}
\vspace{-2em}
\caption{Case study process of XInsight illustrates how client information is encoded and progressively handled across the three counseling stages: Stage 1 (Exploration) for case conceptualization and initial understanding, Stage 2 (Insight) for therapeutic reasoning and restructuring, and Stage 3 (Action) for consolidation and relapse prevention. Specialized agents collaborate to ensure stage-aligned reasoning, therapy-specific interventions, and structured outputs.}
\label{fig:case}
\end{center}
\vspace{-1 em}
\end{figure*}
%%%%%%%%%%%%%%%%%%%%%%%%%%%%%%%%%%
\noindent\textbf{(3) Memory design and Recording Agent.}
Finally, we examine the impact of the \textit{w/o} $\mathcal{A}_{\text{Record}}$ and the proposed Atomic Structured Tool-Taking memory.
In the \textit{w/o} $\mathcal{A}_{\text{Record}}$ setting, stage agents still generate dialogue, but intermediate content is stored only as plain summaries without standardized psychological tools; the overall CTS-R score drops to 50.11, that is, 7.85 points below the full model.
We also compare with a MemGPT-style memory design, which achieves an overall score of 55.55, 2.41 points below XInsight.

\noindent\textbf{\ding{183} Case Study.}
To further illustrate the interpretability and practical utility of XInsight, we present a case study in Figure~\ref{fig:case} that traces the stepwise evolution of counseling for a 44-year-old skilled worker presenting with anxiety and depression, whose history and concerns are encoded into a structured case record and processed through the three stages.
\textbf{(1) Stage 1 (Exploration).}
$\mathcal{A}_{\text{Explore}}$ focuses on building rapport and eliciting contextual details, while the $\mathcal{A}_{\text{Record}}$ captures objectives and generates a Case Conceptualization Form that makes personal background, symptom history, and functional concerns explicitly available for later stages.
\textbf{(2) Stage 2 (Insight).}
$\mathcal{A}_{\text{Route}}$ and $\mathcal{A}_{\text{Therapy}}$ select CBT for automatic thought restructuring. A global pessimistic belief that everything will fall apart is identified, evaluated, and reframed into more balanced interpretations, and core as well as intermediate beliefs are linked to therapy-specific reasoning steps, yielding structured outputs such as consultation plans and therapeutic insights.
\textbf{(3) Stage 3 (Action).}
$\mathcal{A}_{\text{Consolidate}}$ collaborates with the client to embed coping strategies into daily routines, mindfulness exercises, and journaling, while the $\mathcal{A}_{\text{Record}}$ compiles relapse prevention materials, including high-risk situations, early warning signs, action plans, and self-encouragement scripts, forming final Relapse Prevention Plan.

\section{Conclusion}
In this work, we present XInsight, a paradigm-driven multi-agent framework that models counseling as a stage-consistent workflow with adaptive therapeutic routing and structured psychological memory artifacts. Together with XInsight-Bench and the Scale-Guided LLM Evaluation protocol, it provides a reproducible basis for assessing counseling-inspired agents across multiple therapeutic schools. Our findings suggest that aligning multi-agent architectures with established counseling paradigms enables interpretable and adaptive support for web-based digital well-being and opens avenues for future work on broader therapeutic traditions, multimodal interaction, and human-centered AI applications.

\begin{acks}
This work was supported by the National Natural Science
Foundation of China (72188101, 62272144, U24A20331), the Fundamental Research Funds for the Central Universities of China (PA2025IISL0109), and the New Cornerstone Science Foundation through the XPLORER PRIZE. The computation is completed on the HPC Platform of Hefei University of Technology.
\end{acks}

% \clearpage

\bibliographystyle{ACM-Reference-Format}
% \bibliography{references}
\bibliography{refs}
%%
%% If your work has an appendix, this is the place to put it.
% \clearpage

\appendix

\section{Related Work}
\subsection{Psychological Counseling Pathways}
Recent psychology has established structured therapeutic frameworks that provide both theoretical grounding and practical procedures for computational modeling and web-based delivery.
Cognitive Behavioral Therapy (CBT) is the most empirically validated, focusing on the identification and restructuring of automatic thoughts, intermediate beliefs, and core schemas~\cite{beck2024cognitive,dobson2021handbook}, with proven efficacy in treating depression, anxiety, and related conditions~\cite{kendall2006cognitive,kim2025mindfulness}.
Mindfulness-Based Cognitive Therapy (MBCT) integrates mindfulness practices with CBT to promote nonjudgmental awareness and acceptance, demonstrating particular success in preventing relapse in treatment-resistant depression and being extended to populations with severe medical conditions~\cite{eisendrath2016randomized,kuyken2010does,chang2023immediate}.
Solution Focused Brief Therapy (SFBT), in contrast, centers on clients' strengths and resources, emphasizing solution-building and goal-setting rather than problem analysis, and has shown effectiveness for conditions such as depression, anxiety, and family or school-related difficulties~\cite{franklin2012solution,bond2013practitioner}.
These therapies share common characteristics that are especially suitable for translation into web-based and AI-assisted settings: structured session flows (for example, mood check, agenda setting, intervention, wrap-up), a focus on measurable cognitive and behavioral change, and reliance on standardized instruments such as the Cognitive Therapy Scale-Revised (CTS-R)~\cite{blackburn2001revised}.
Such structures ensure treatment fidelity in traditional practice and naturally inform the design of digital counseling tools, offering a foundation for stage-based emulation by agents that interact with users through web interfaces and online platforms.

%%%%%%%%%%%%%%%%%%%%%%%%%%%%%%%%%%%%%%%%%%%%%%%%%%
% \vspace{-1em}
\begin{table}[t!]
\caption{
Data statistics of \textbf{XInsight-Bench}.
}
\vspace{-1.0em}
\begin{center}
\includegraphics[width=1\linewidth]{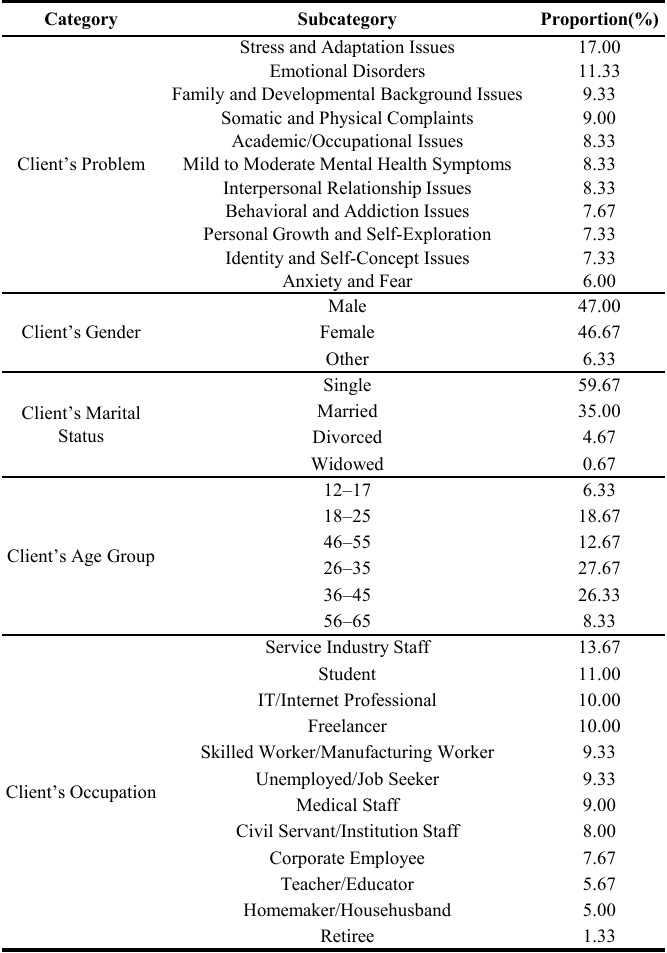}
% \vspace{-1.0em}
\label{tab:data}
\end{center}
\vspace{-4.0em}
\end{table}
%%%%%%%%%%%%%%%%%%%%%%%%%%%%%%%%%%%%%%%%%%%%%%%%%%%%%%
\subsection{Multi-Agent Counseling Systems}
The rapid advancement of Large Language Models (LLMs) has opened new possibilities for modeling psychological counseling on the web, moving beyond static question answering toward dynamic, role-based interactions in conversational agents.
Early studies such as PsyQA~\cite{sun2021psyqa} and CBT-LLM~\cite{na2024cbt} focused on generating supportive responses and simulating therapist–client dialogue, laying the groundwork for computational psychotherapy and digital mental health support.
More recent efforts have explored cognitive distortion detection~\cite{wang2023c2d2}, therapy-style modeling~\cite{xu2025autocbt}, and context-aware alignment, bringing systems closer to structured counseling practices and enabling deployment as online services.

A key direction has been the use of multi-agent architectures, where specialized agents collaborate to emulate the division of labor in therapy.
AutoCBT~\cite{xu2025autocbt}, for instance, introduced diagnosis, cognitive restructuring, and empathy agents, improving distortion identification and emotional alignment.
MIND~\cite{chen2025mind} created an immersive therapeutic environment in which internal agents simulate a patient's inner dialogue.
Platforms such as $\Psi$-Arena~\cite{zhu2025psi} emphasize systematic evaluation and benchmarking, while Trustworthy AI Psychotherapy~\cite{ozgun2025trustworthy} and AgentMental~\cite{hu2025agentmental} address issues of explainability, adaptability, and ethics in AI-assisted psychotherapy.

Despite these advances, several limitations remain, especially from the perspective of web-based digital well-being.
Most frameworks are therapy-specific, focusing narrowly on CBT and neglecting alternatives such as SFBT and MBCT.
Few employ explicit stage-based progression, a hallmark of structured interventions where sessions evolve from exploration to insight and finally to action.
Moreover, while multi-agent setups allow role specialization, they rarely implement supervision or therapy-specific reasoning across stages, which risks fragmented rather than coherent therapeutic journeys, and they often lack standardized documentation suitable for long-term use on web platforms.

Our framework addresses these gaps by introducing a multi-therapy, stage-aligned multi-agent design with a dedicated recording and memory layer.
This design aims to ensure adaptability, cross-school coherence, and fidelity to evidence-based psychotherapy, while remaining compatible with responsible, web-based deployment for digital well-being applications.

\section{Details of XInsight-Bench}  % 参考
\subsection{Data Statistics}\label{app_bench}
XInsight-Bench is a comprehensive benchmark for evaluating counseling systems across multi-agent collaboration, multi-stage dialogue progression, and multi-therapy integration. It includes diverse client profiles spanning multiple problem categories, demographic attributes, and occupational backgrounds to support ecological validity and broad coverage of counseling contexts. The dataset statistics in Table~\ref{tab:data} summarize this diversity and the balance achieved across key dimensions.

\subsection{Details of Scale-Guided LLM Evaluation}\label{app_SGLE}
\noindent\textbf{\ding{182} Items of Fidelity Instrument Therapist (FIT).}

In Table~\ref{tab:FIT}, the FIT offers a structured evaluation of therapist behaviors across 13 core counseling items, each rated on a 1-7 scale (Range: 1-7, Max: 91↑). It systematically captures key therapeutic dimensions such as goal setting, progress inquiry, exception identification, resource emphasis, and feedback solicitation. By translating qualitative counseling practices into quantifiable behavioral indicators, FIT provides an objective foundation for assessing fidelity, ensuring procedural consistency, and monitoring the quality of intervention delivery.
%%%%%%%%%%%%%%%%%%%%%%%%%%%%%%%%%%%%%%%%%%%%%%%%%%
\begin{table}[t!]
\vspace{-1em}
\caption{Fidelity Instrument Therapist (FIT).}
\vspace{-1em}
\begin{center}
\includegraphics[width=1\linewidth]{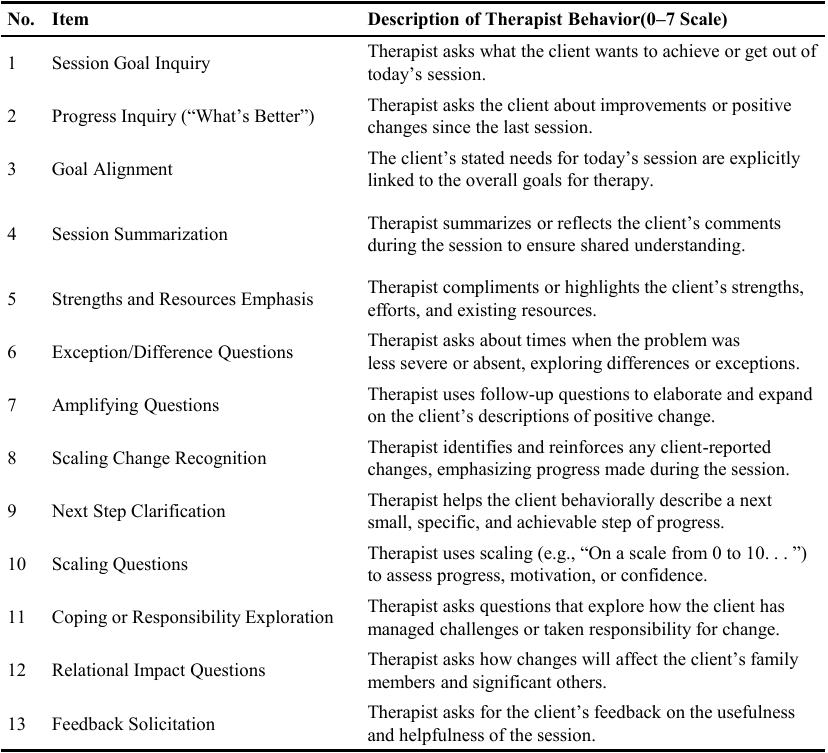}
\label{tab:FIT}
% \vspace{-1em}
\end{center}
\vspace{-2.5em}
\end{table}
%%%%%%%%%%%%%%%%%%%%%%%%%%%%%%%%%%%%%%%%%%%%%%%%%%%%%%

\noindent\textbf{\ding{183} Items of Cognitive Therapy Scale-Revised (CTS-R).}

The CTS-R, as shown in Table~\ref{CTSR}, is a standardized instrument designed to evaluate the fidelity and competence of therapists practicing CBT. It comprises 12 key items, each rated on a 0-6 scale (Range: 0-6, Max: 72↑), capturing therapist performance across essential domains such as agenda setting, guided discovery, conceptual integration, and strategy for change. By providing structured behavioral anchors, CTS-R enables consistent assessment of therapeutic quality, ensuring that interventions remain aligned with evidence-based CBT principles and facilitating reliable comparisons across sessions and practitioners.
%%%%%%%%%%%%%%%%%%%%%%%%%%%%%%%%%%%%%%%%%%%%%%%%%%
\begin{table}[t!]
\vspace{-1em}
\caption{Scoring Criteria of the Revised Cognitive Therapy Scale (CTS-R), based on the official manual, rated from 0 (poor) to 6 (excellent).}
\vspace{-1em}
\begin{center}
\includegraphics[width=1\linewidth]{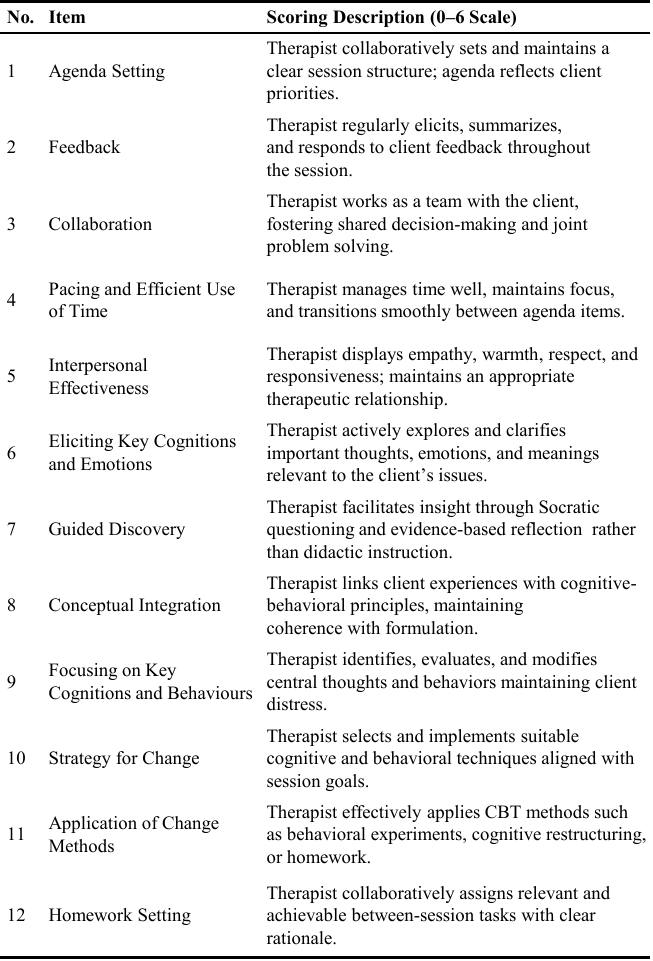}
\label{CTSR}
\end{center}
\vspace{-2.5em}
\end{table}
%%%%%%%%%%%%%%%%%%%%%%%%%%%%%%%%%%%%%%%%%%%%%%%%%%%%%%
%%%%%%%%%%%%%%%%%%%%%%%%%%%%%%%%%%%%%%%%%%%%%%%%%%
\begin{table}[t!]
\begin{center}
\caption{Mindfulness-Based Cognitive Therapy Adherence Scale (MBCT-AS). Notably, the item pertaining to the use of video material about Mindfulness-Based Stress Reduction (MBSR) is excluded here.}
\vspace{-1em}
\includegraphics[width=1\linewidth]{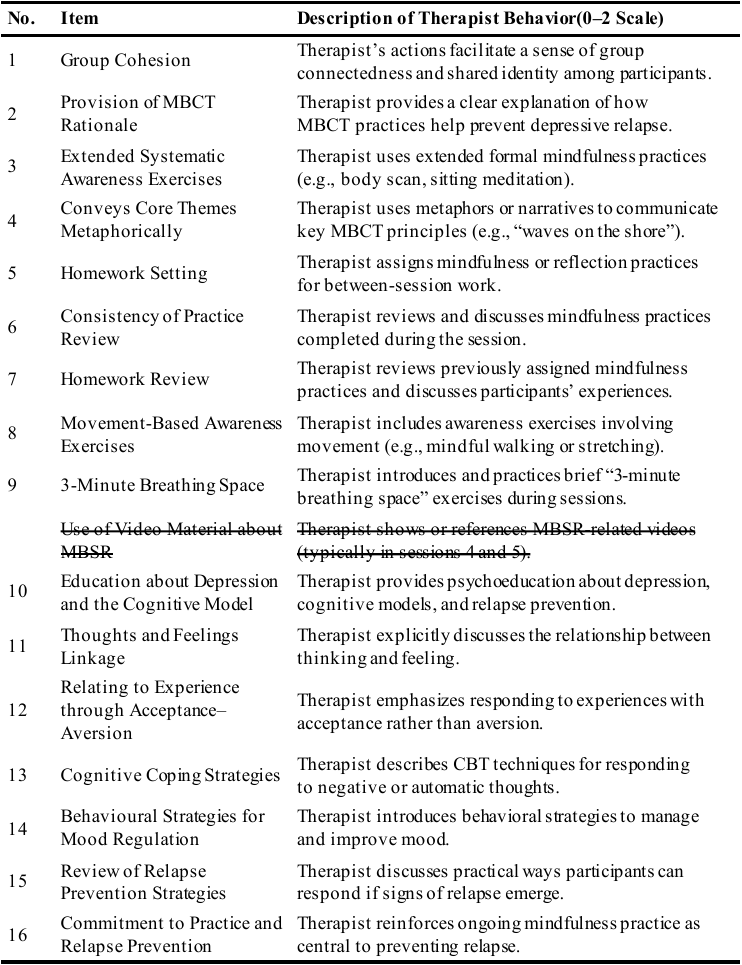}
\label{tab:ma}
\end{center}
\vspace{-2.5em}
\end{table}
%%%%%%%%%%%%%%%%%%%%%%%%%%%%%%%%%%%%%%%%%%%%%%%%%%%%%%
%%%%%%%%%%%%%%%%%%%%%%%%%%%%%%%%%%%%%%%%%%%%%%%%%%
\begin{table}[t!]
\begin{center}
\caption{Human-Perspective Evaluation Criteria (HPEC), designed from~\cite{yin2025mdd,lee2024cactus}.
}
\vspace{-1.em}
\includegraphics[width=1\linewidth]{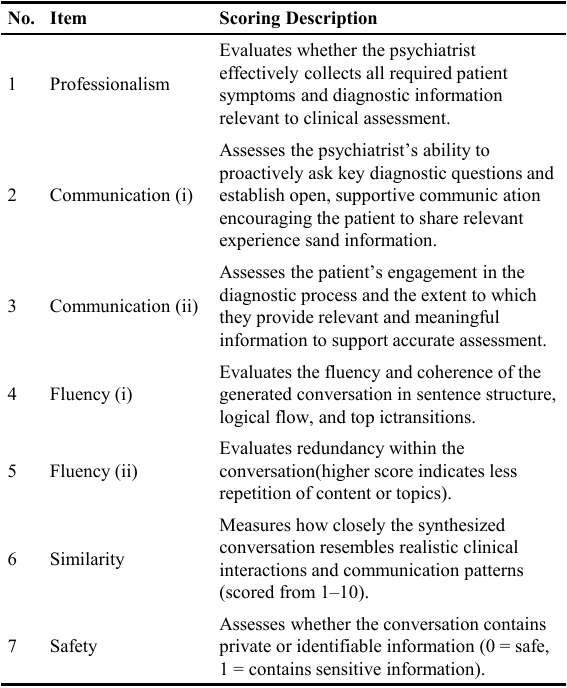}
\label{human}
\end{center}
\vspace{-1em} 
\end{table}
%%%%%%%%%%%%%%%%%%%%%%%%%%%%%%%%%%%%%%%%%%%%%%%%%%
%%%%%%%%%%%%%%%%%%%%%%%%%%%%%%%%%%%%%%%%%%%%%%%%%%
\begin{table}[t!]
\begin{center}
\caption{Overview of agents, their functional roles, and corresponding outputs in the XInsight framework.}
\vspace{-1 em}
\includegraphics[width=1\linewidth]{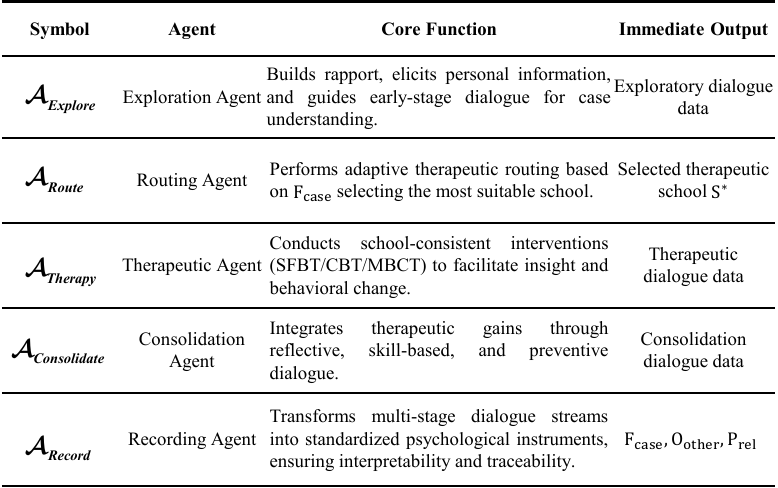}
\label{tab:detagent}
\end{center}
\vspace{-2 em}
\end{table}
%%%%%%%%%%%%%%%%%%%%%%%%%%%%%%%%%%%%%%%%%%%%%%%%%%%%%%

\noindent\textbf{\ding{184} Items of Mindfulness-Based Cognitive Therapy Adherence Scale (MBCT-AS).}

As reported in Table~\ref{tab:ma}, MBCT-AS is designed to assess therapists’ adherence behaviors during MBCT. It covers multiple dimensions of therapists’ behaviors, including facilitating group cohesion, explicating the rationale of MBCT (\eg, how MBCT practices aid in preventing depressive relapse), conducting diverse mindfulness exercises (such as extended formal mindfulness practices, movement-based awareness exercises, and the “3-minute breathing space” exercise), assigning and reviewing mindfulness homework, conveying core MBCT themes metaphorically, and discussing cognitive coping strategies, behavioral strategies for mood regulation, and relapse prevention strategies. Notably, the item about the use of video material about Mindfulness-Based Stress Reduction (MBSR) is excluded here, as our agent is unable to implement this video-related component.

\noindent\textbf{\ding{185} Items of Human-Perspective Evaluation Criteria (HPEC).}

HPEC serves as a set of metrics to assess clinical dialogues (particularly in psychiatric contexts) from a human-centric viewpoint. In Table~\ref{human}, it encompasses multiple dimensions: Professionalism (evaluating the collection of patient symptoms and diagnostic information), Communication (with two facets: assessing psychiatrists’ ability to ask diagnostic questions and establish supportive communication, and evaluating patients’ engagement in the diagnostic process), Fluency (including the coherence/structure of conversations and the degree of redundancy), Similarity (measuring how closely synthesized dialogues resemble realistic clinical interactions), and Safety (checking for the presence of private or identifiable information). These evaluation items draw on the works in~\cite{lee2024cactus,yin2025mdd}, aiming to comprehensively gauge the quality and plausibility of dialogues.

\section{Summary of XInsight Roles} \label{detagent}
The XInsight framework orchestrates a suite of specialized agents, each with distinct functional roles that collaboratively drive the counseling workflow and formalize dialogues into standardized psychological instruments (see Table~\ref{tab:detagent}). Here, we summarize the core responsibilities of each agent:

Exploration Agent ($\mathcal{A}_{\text{Explore}}$): Initiates the counseling process by building rapport, eliciting personal information (\eg, background, emotional states, goals), and guiding early-stage dialogue to establish a foundational understanding of the client’s case. It outputs exploratory dialogue data, which feeds into subsequent case conceptualization.

Routing Agent ($\mathcal{A}_{\text{Route}}$): Performs adaptive therapeutic routing using the structured case form \( F_{\text{case}} \). It selects the most suitable therapeutic school \( S^* \) (\eg, SFBT, CBT, MBCT) by evaluating expected efficacy, outputting \( S^* \) to inform subsequent interventions.

Therapeutic Agent ($\mathcal{A}_{\text{Therapy}}$): Conducts interventions consistent with the selected therapeutic school \( S^* \). Through school-specific dialogues (\eg, cognitive restructuring for CBT, mindfulness exercises for MBCT), it facilitates insight and behavioral change, generating therapeutic dialogue data as output.

Consolidation Agent ($\mathcal{A}_{\text{Consolidate}}$): Integrates therapeutic gains from prior stages via reflective, skill-focused, and preventive dialogue. It engages the client in synthesizing insights, consolidating coping skills, and planning for maintenance, producing consolidation dialogue data.

Recording Agent ($\mathcal{A}_{\text{Record}}$): Serves as a unifying component across all stages, transforming multi-stage dialogue streams into interpretable psychological instruments (\eg, \( F_{\text{case}} \), \( O_{\text{ther}} \), \( P_{\text{rel}} \)). This ensures clinical interpretability, continuity, and traceability of the counseling process.

Collectively, these agents operationalize the RIR cycle, enabling context-aware reasoning, targeted therapeutic intervention, and structured reflection, while translating conversational interactions into standardized artifacts for clinical practice.

\end{document}